\documentclass[useAMS, usenatbib, fleqn]{mn2e}
\pdfoutput=1
\usepackage{aecompl}
\usepackage{amsmath}
\usepackage{amssymb}
\usepackage{color}
\usepackage{graphicx}
\usepackage{multirow}

\usepackage{url}
\usepackage{hyperref}
\hypersetup{colorlinks=true,citecolor=blue,linkcolor=blue,filecolor=blue,urlcolor=blue}

\newcommand*\diff{\mathop{}\!\mathrm{d}}

\title[Dust Formation in Milky Way-like Galaxies]{Dust Formation in Milky Way-like Galaxies}
\author[R.~McKinnon, P.~Torrey, and M.~Vogelsberger]
    {Ryan McKinnon$^{1}$\thanks{E-mail: ryanmck@mit.edu},
     Paul Torrey$^{1,2}$, and Mark Vogelsberger$^{1}$ \\
     $^{1}$Department of Physics and Kavli Institute for Astrophysics and Space Research,
           Massachusetts Institute of Technology, \\
           Cambridge, MA 02139, USA  \\
     $^{2}$TAPIR, California Institute of Technology, Pasadena, CA 91125, USA
    }
\begin{document}

\date{Accepted 2016 January 28. Received 2016 January 28; in original form 2015 May 18}

\pagerange{\pageref{firstpage}--\pageref{lastpage}} \pubyear{2016}

\maketitle

\label{firstpage}

\begin{abstract}
We introduce a dust model for cosmological simulations implemented in the
moving-mesh code \textsc{arepo} and present a suite of cosmological
hydrodynamical zoom-in simulations to study dust formation within galactic
haloes.  Our model accounts for the stellar production of dust, accretion of
gas-phase metals onto existing grains, destruction of dust through local
supernova activity, and dust driven by winds from star-forming regions.  We
find that accurate stellar and active galactic nuclei feedback is needed to
reproduce the observed dust-metallicity relation and that dust growth largely
dominates dust destruction.  Our simulations predict a dust content of the
interstellar medium which is consistent with observed scaling relations at $z =
0$, including scalings between dust-to-gas ratio and metallicity, dust mass and
gas mass, dust-to-gas ratio and stellar mass, and dust-to-stellar mass ratio
and gas fraction.  We find that roughly two-thirds of dust at $z = 0$
originated from Type II supernovae, with the contribution from asymptotic giant
branch stars below 20 per cent for $z \gtrsim 5$.  While our suite of Milky
Way-sized galaxies forms dust in good agreement with a number of key
observables, it predicts a high dust-to-metal ratio in the circumgalactic
medium, which motivates a more realistic treatment of thermal sputtering of
grains and dust cooling channels.
\end{abstract}


\section{Introduction}

Dust in the interstellar medium (ISM) exists alongside gas-phase metals
and alters the dynamic and spectroscopic properties of galaxies
\citep{Calzetti1994, Silva1998, Dey1999, Calzetti2000, Netzer2007, Spoon2007,
Melbourne2012}.  The surfaces of dust grains play host to a range
of chemical reactions that subsequently influence the behaviour of the ISM and
impact star formation \citep{Hollenbach1971, Mathis1990, Li2001, Draine2003}.
Additionally, observations suggest that dust is a significant contributor of
metals in the circumgalactic medium \citep[CGM;][]{Bouche2007, Peeples2014,
Peek2015}.  Understanding the life cycle of dust grains, including their
production in asymptotic giant branch (AGB) stars and supernovae
\citep[SNe;][]{Gehrz1989, Todini2001, Nozawa2003, Ferrarotti2006, Nozawa2007,
Zhukovska2008, Nanni2013, Schneider2014}, growth via accretion of gas particles
in the ISM and coagulation with other dust particles \citep{Draine1990,
Dominik1997, Dwek1998, Hirashita2011}, destruction via thermal sputtering,
collisions with other dust grains, and SN shocks \citep{Draine1979a,
Draine1979b, McKee1989, Jones1996, Bianchi2005, Yamasawa2011}, and other
physical processes, is important in accurately modelling dust evolution.

Even at high redshift, galaxies can form substantial masses of dust.
Far-infrared and submillimeter observations show that some dust-rich radio
galaxies and quasars out to $z \sim 7$ have dust masses greater than $10^7 \,
\textrm{M}_\odot$ \citep{Fan2003, Bertoldi2003, Hughes1997, Venemans2012,
Casey2014, Riechers2014, Watson2015}.  Two notable examples are SDSS
J1148+5251, a $z = 6.4$ quasar with an inferred dust mass of $\left(
3.4^{+1.38}_{-1.54} \right) \times 10^8 \, \text{M}_\odot$ \citep{Valiante2009,
Valiante2011}, and A1689-zD1, a $z = 7.5 \pm 0.2$ galaxy with a dust mass of $4
\times 10^{7} \, \text{M}_\odot$ and a dust-to-gas ratio comparable to the
Milky Way value \citep{Watson2015}.  To produce such large dust masses at high
redshift, average SNe must yield roughly $1 \, \text{M}_\odot$ of dust, a
quantity larger than the amount of dust SNe have been observed to condense
\citep{Todini2001, Sugerman2006, Dwek2007, Lau2015}.  However, these dusty
examples may not be representative of typical high-redshift galaxies.  There is
recent evidence of actively star-forming galaxies at $z > 6.5$ with little dust
obscuration \citep{Walter2012, Kanekar2013, Ouchi2013}, and observations
suggest that dust extinction is generally decreased in low luminosity and high
redshift systems \citep{Bouwens2012}.  Our current understanding of dust formation
at high redshift is therefore still quite incomplete.

Significant variation in dust properties also exists at low redshift.
The Galactic dust-to-gas ratio is roughly $10^{-2}$ \citep{Gilmore1989,
Sodroski1997, Zubko2004} and several times larger than the values observed for
the Large and Small Magellanic Clouds \citep{Pei1992, Gordon2014,
RomanDuval2014, Tchernyshyov2015}.  In contrast, the metal-poor local dwarf
galaxy I Zwicky 18 has been estimated to have a dust mass of no more than $1800
\, \text{M}_\odot$ and a corresponding dust-to-gas ratio in the range of
$(3.2-13) \times 10^{-6}$, several orders of magnitude below the typical values
expected for larger systems \citep{Fisher2014}.  Additionally, observations and
models of the ISM indicate that the nature of dust can differ among individual
chemical species \citep{Wilms2000, Kimura2003, Jenkins2009}.

The current sample of galaxies with reliable dust estimates has grown in recent
years, driven by programs like SINGS \citep{Kennicutt2003, Draine2007}, the
Herschel Reference Survey \citep{Boselli2010}, the Herschel ATLAS
\citep{Eales2010}, KINGFISH \citep{Kennicutt2011}, and the Dwarf Galaxy Survey
\citep{Madden2013}.  Various trends and scaling relations involving dust and
host galaxy properties have been observed from this data.  There is a positive
correlation between dust-to-gas ratio and metallicity \citep{Vladilo1998,
Draine2007, Galametz2011, RemyRuyer2014, Zahid2014}.  However, the detailed
behaviour of the dust-metallicity relation is unclear.  Observations indicate
that the dust-to-gas ratio is reduced at low metallicity, possibly due to
effects in the interstellar radiation field that limit dust growth and enhance
destruction processes \citep{RemyRuyer2014}.  Even over small metallicity
ranges, there is pronounced scatter in the dust-to-gas ratio.

The observed dust-metallicity relation implies a dust-to-metal ratio that is
fairly constant across a range of galaxy morphologies and histories.  However,
the nature of the dust-to-metal ratio at high redshift and low metallicity is
uncertain.  Recent work using gamma ray bursts has yielded dust-to-metal ratios
fairly consistent with the Local Group, even in low metallicity systems
\citep{Zafar2013, Sparre2014}.  This would require SNe to be efficient
producers of dust or grains to grow rapidly in the ISM \citep{Mattsson2014a}.
Separate analysis of gamma ray burst damped Lyman-alpha absorbers suggests a
non-universal dust-to-metal ratio, with low metallicity environments producing
dust less efficiently than spiral galaxies \citep{DeCia2013}.  Even in the
Milky Way, observations of strong gas-phase depletion \citep{Roche1985,
Sembach1996, Jenkins2004, Jenkins2009} contrast the expectation that dust
destruction outpaces stellar injection of dust \citep{Barlow1978, Draine1979a,
McKee1989, Dwek1980, Jones1994, Jones1996}.  Understanding the balance between
gas-phase metals and dust in Milky Way-sized systems requires more work.

A number of other empirical scaling relations have emerged, tying observed dust
masses to various galactic properties.  These include relations between
dust-to-stellar mass ratio and gas fraction \citep{Cortese2012},
dust-to-stellar mass ratio and redshift \citep{Dunne2011}, dust extinction and
stellar mass \citep{Zahid2014}, dust mass and gas mass \citep{Corbelli2012},
dust mass and star formation rate \citep{DaCunha2010}, and dust surface density
and radial distance \citep{Menard2010, Pappalardo2012}.  Observational data have
also yielded initial estimates of the dust mass function for low and high
redshift \citep{Dunne2000, Dunne2011}.  These scaling relations provide
constraints on galaxy formation models that include a treatment of dust.

A variety of numerical models have been used in previous work to better
understand how dust evolves in a galaxy.  These include one- and two-zone
models \citep{Dwek1998, Lisenfeld1998, Hirashita2002, Inoue2003a, Morgan2003,
Calura2008, Valiante2009, Gall2011a, Yamasawa2011, Asano2013a, Zhukovska2014,
Feldmann2015}, semi-analytic methods \citep{Somerville2012, Mancini2015}, and
more recently first smoothed-particle hydrodynamical simulations resolving
local dust variations \citep{Bekki2013, Bekki2015a}.  These models include
processes like the formation of dust during stellar evolution, dust growth and
destruction in the ISM, radiation field effects, and dust-enhanced molecular
formation.  Many of these one-zone models have addressed the formation of dusty
high-redshift quasars and have indicated that if dust is unable to condense
efficiently in SNe, a high star formation or accretion efficiency is needed to
produce large dust masses at $z > 5$ \citep{Morgan2003, Michalowski2010,
Gall2011b, Gall2011c, Michalowski2015}.

Numerical dust models have also focused on the evolution of the dust-to-gas and
dust-to-metal ratios, often used in observations to characterise galaxies.  The
dust-metallicity relation is reproduced in many one-zone models
\citep{Issa1990, Lisenfeld1998}.  These models predict a dust-to-metal ratio
that is independent of metallicity in galaxies where SNe are the primary
producers of dust \citep{Morgan2003} and whose radial gradient in galactic
discs can be used to estimate interstellar dust growth \citep{Mattsson2012}.
The evolution of the dust-to-metal ratio may vary significantly with time, with
the dust-to-metal ratio at $z \gtrsim 2$ possibly just 20 per cent of the
present value \citep{Inoue2003a}.  Interestingly, galaxies are predicted to
switch from low to high dust-to-metal ratio when crossing a critical
metallicity threshold that enables efficient ISM dust growth
\citep{Zhukovska2008, Inoue2011, Asano2013a, Feldmann2015}.  Previous one-zone
dust models applied to the formation of a single galaxy show present day
dust-to-metal ratios of roughly $0.5$ \citep{Dwek1998} and $0.9$
\citep{Calura2008}, with results significantly dependent on the strength of
dust growth and destruction mechanisms in the ISM.  These models also find
depletion roughly constant across all chemical species.  However, the biggest
weakness of these one-zone models is their lack of spatial resolution, limiting
their ability to make predictions about the distribution of dust within a
galaxy.

Cosmological simulations provide a better means to understand how dusty systems
can form at high redshift and how their dust content, both in the ISM and CGM,
evolves in time.  Additionally, simulations of full cosmological volumes can
provide the sample size of galaxies needed to corroborate observed scaling
relations involving dust.  Recent cosmological hydrodynamical simulations have
suggested that heavily dust-attenuated galaxies can form even at $z \sim 7$,
with ultraviolet and optical colors consistent with a model having a
dust-to-metal ratio of $0.4$ \citep{Kimm2013}.  Motivated by the observed
reddening of background quasars by foreground galaxies in Sloan Digital Sky
Survey (SDSS) data \citep{Menard2010}, smoothed-particle hydrodynamical
simulations found that half of this reddening signal is attributable to dust
more than $100 \, h^{-1} \, \text{kpc}$ from the closest massive galaxy
\citep{Zu2011}.  Radial gradients of the dust-to-gas ratio in galactic discs
appear to be steeper for larger galaxies \citep{Bekki2015a}.  Even in
simulations where dust is not directly treated, radiative transfer
post-processing can be used to infer dust extinction \citep{Jonsson2006,
Narayanan2010, Hayward2013, Yajima2014}.  The dynamics of grain sputtering in
SN shocks can also be studied using tracer particles \citep{Silvia2010}.
While some studies have investigated the impact of feedback mechanisms, such as
winds driven by SNe, on the cosmological evolution of dust \citep{Zu2011,
Bekki2015a}, much work still remains.

In this work, we incorporate essential dust physics into a moving-mesh
simulation code and, for the first time in studies of dust, use a large sample
of zoom-in cosmological initial conditions.  Compared to previous dust models
using one-zone methods and idealised initial conditions, our approach has the
ability to resolve the structure of dust within a galaxy and the impact that
local feedback processes have on dust evolution.  While some smoothed-particle
hydrodynamical simulations have treated dust without using zoom-in cosmological
initial conditions, our highest resolution cosmological run offers improved
mass resolution.

This paper is structured as follows.  In Section~\ref{SEC:methods}, we
discuss the theory and numerical implementation of our dust model in the
context of a larger galaxy formation framework.  Details of the cosmological
initial conditions for our simulations are provided in
Section~\ref{SEC:cosmological_runs}.  We analyse the importance of physical
feedback processes, individual dust model components, and variations in
simulation initial conditions in Section~\ref{SEC:results}.  Finally, in
Section~\ref{SEC:discussion}, we summarise our results and discuss the
implications of our findings on future observations.

\section{Methods}\label{SEC:methods}

We employ cosmological simulations to track the evolution of dust and its effect
on galaxy dynamics.  We briefly review the core components of the galaxy
formation model currently implemented in the moving-mesh code \textsc{arepo}
\citep{Springel2010}, which have been detailed in prior work
\citep{Vogelsberger2013} and used for various cosmological studies
\citep{Vogelsberger2012, Genel2014, Torrey2014, Vogelsberger2014a,
Vogelsberger2014b}, and then describe the new dust physics we have added.
\textsc{arepo} uses a dynamic Voronoi tessellation to solve the equations of
ideal hydrodynamics with a finite-volume method.  A second-order Godunov scheme
is used in conjunction with an exact Riemann solver to compute fluxes between
cells.  Additionally, gravitational and collisionless physics have been
implemented using methods similar to the TreePM scheme in the smoothed-particle
hydrodynamics code \textsc{gadget} \citep{Springel2005b}.

\subsection{Galaxy Formation and Feedback Mechanisms}\label{SEC:galaxy_formation}

The current galaxy formation model in \textsc{arepo} accounts for a number of
physical processes, including primordial and metal-line gas cooling, stellar
evolution and subsequent chemical enrichment of the ISM, black hole formation,
and stellar and active galactic nuclei (AGN) feedback, which together yield a
galaxy stellar mass function in good agreement with observations over $0 < z <
3$ \citep{Torrey2014, Genel2014}.  The star formation prescription
stochastically creates star particles in dense regions of ISM gas, with masses
distributed according to a \citet{Chabrier2003} initial mass function (IMF).  As
stars evolve off of the main sequence, mass is recycled to the neighboring ISM.
This chemical enrichment routine follows nine elements (H, He, C, N, O, Ne, Mg,
Si, and Fe).  We adopt elemental mass yields and recycling fractions for AGB
stars from \citet{Karakas2010}, SNe II from \citet{Portinari1998}, and SNe Ia
from \citet{Thielemann2003}.  We employ a stellar lifetime function from
\citet{Portinari1998}.  Galactic outflows are modelled using wind particles
launched from star-forming ISM gas and temporarily decoupled from hydrodynamics
to emulate a galactic fountain driven by stellar feedback \citep{Springel2003}.
We include the minor feedback modifications detailed in \cite{Marinacci2014},
which alter radio-mode AGN feedback and adopt warm galactic winds.

\subsection{Dust Evolution}\label{SEC:dust_model}

A wide variety of dust models have been used in recent galaxy formation
simulations, though essentially all models track both the formation of dust
during stellar evolution and its subsequent evolution in the ISM.  The most
simplistic of these models take a one-zone approach, solving a set of
coupled ordinary differential equations for the time evolution of the total
mass of gas, metals, and dust within a galaxy \citep{Hirashita2002, Inoue2003a,
Morgan2003, Calura2008, Valiante2009, Gall2011a, Yamasawa2011, Asano2013a,
Zhukovska2014, Feldmann2015}.  Other work assumes a two-component galaxy model,
consisting of a bulge and disc region and allowing for local dust properties to
be studied as a function of radial distance from the galactic center
\citep{Dwek1998}.  More recently, \citet{Bekki2013, Bekki2015a} has performed
smoothed-particle hydrodynamical simulations that treat dust locally and offer
improved spatial resolution.

In this work, we focus on the dominant dust production and evolution
mechanisms.  We leave additional processes, such as the catalysis of molecular
hydrogen on dust grains and the effect of interstellar radiation fields, to
future efforts.  We track the mass of dust in each chemical species within
individual gas cells.  Dust is passively advected between gas cells when
solving the hydrodynamic equations in each time step, in essence adopting a
two-fluid approach with dust fully coupled to gas.  However, dust is not
implemented as a strictly passive tracer: as shown below, dust impacts
metal-line cooling and in turn star formation and gas dynamics.  Alternative
treatments, including ``live'' dust particles decoupled from gas and subject to
frictional or radiative forces \citep{Kwok1975, Draine1979b, Barsella1989,
Theis2004, Bekki2015b}, may be pursued in future work.

\subsubsection{Dust Production via Stellar Evolution}\label{SEC:dust_production}

During the stellar evolution process described in
Section~\ref{SEC:galaxy_formation}, a certain amount of the mass $\Delta M_{i}$
of species $i$ evolved by stars and returned to neighboring gas cells in some
time step is assumed to condense into dust, with the exception of H,
He, N, and Ne.  The remaining metal mass exists in the gas phase.  In the
framework below, we follow the approach used by \citet{Dwek1998} and other
subsequent works that track individual chemical elements.  Most notably, we
adopt a different functional form for the amount of dust produced during mass
return from AGB stars than from SNe.  Additionally, we make a distinction
between AGB stars with $\text{C}/\text{O} > 1$ in their stellar envelope, which
are expected to produce carbonaceous solids (e.g.\ graphite or amorphous
carbon), and those with $\text{C}/\text{O} < 1$, which are thought to form
primarily silicate dust \citep{Draine1990}.  In essence, the equations below
describe elemental mass yields for dust, computed as a function of the yields
for overall metals returned by stars.

For AGB stars with a carbon-to-oxygen ratio of $\text{C}/\text{O} > 1$ in their
returned mass, the amount of species $i$ dust produced is given by
\begin{equation}
\Delta M_{i,\text{dust}} =
\begin{cases}
\delta_\text{C}^\text{AGB,$\text{C}/\text{O}>1$} (\Delta M_{\text{C}} - 0.75 \, \Delta M_\text{O}) \; &\text{if $i = \text{C}$}\\
0 \; &\text{else},
\end{cases}
\end{equation}
where $\delta_\text{C}^\text{AGB,$\text{C}/\text{O}>1$}$ is the carbon condensation
efficiency for AGB stars with $\text{C}/\text{O} > 1$, discussed below in more detail.
Similarly, for AGB mass return with $\text{C}/\text{O} < 1$, the mass of
species $i$ dust formed is
\begin{equation}
\Delta M_{i,\text{dust}} =
\begin{cases}
0 \; &\text{if $i = \text{C}$} \\
10 \sum\limits_{j = \text{Mg, Si, Fe}} \delta_{j}^\text{AGB,$\text{C}/\text{O}<1$} \Delta M_{j} / \mu_{j} \; &\text{if $i = \text{O}$} \\
\delta_{i}^\text{AGB,$\text{C}/\text{O}<1$} \Delta M_{i} \; &\text{else}, \\
\end{cases}
\label{EQN:AGB_O_yield}
\end{equation}
where $\mu_{i}$ is the mass in amu and $\delta_{i}^\text{AGB,$\text{C}/\text{O}<1$}$
is the condensation efficiency for species $i$ in AGB stars with $\text{C}/\text{O} < 1$.
Finally, the mass of dust for element $i$ produced via SNe ejecta is
\begin{equation}
\Delta M_{i,\text{dust}} =
\begin{cases}
\delta_\text{C}^\text{SN} \Delta M_\text{C} \; &\text{if $i = \text{C}$} \\
10 \sum\limits_{j = \text{Mg, Si, Fe}} \delta_{j}^\text{SN} \Delta M_{j} / \mu_{j} \; &\text{if $i = \text{O}$} \\
\delta_{i}^\text{SN} \Delta M_{i} \; &\text{else}, \\
\end{cases}
\label{EQN:SNe_dust_yields}
\end{equation}
where the dust condensation efficiency of element $i$ for SNe is
$\delta_{i}^\text{SN}$, which may differ from the corresponding efficiency for
AGB stars.  Equation~(\ref{EQN:SNe_dust_yields}) and its condensation
efficiencies are used for both SNe Ia and II, consistent with all
aforementioned models.  Additionally, we note that the numerical prefactor for
the calculation of $\Delta M_\text{O,dust}$ in Equations~(\ref{EQN:AGB_O_yield})
and (\ref{EQN:SNe_dust_yields}) is slightly reduced from that given in
\citet{Dwek1998}.  We have found that this minor modification is necessary for
oxygen, whose dust production is tied to the ejecta of heavier elements, in
order to avoid the formation of more oxygen dust than total oxygen mass
returned.  We use different stellar nucleosynthesis yields than in
\citet{Dwek1998}, and even though there is considerable uncertainty in the
condensation efficiencies, we demonstrate in Section~\ref{SEC:variations_dust}
that our results are insensitive to moderate changes in these prefactors.  The
range of condensation efficiencies that we explore is described in
Section~\ref{SEC:fiducial}.

Within each gas cell, we separately track the mass of dust produced by AGB
stars, SNe Ia, and SNe II, motivated by a desire to understand the dominant
channels of dust production at high redshift.  Additionally, we refer to carbon
grains as graphite dust, and the dust in remaining species as silicate dust.
While a simplification that fails to distinguish the full diversity of
compounds that comprise dust grains, including polycyclic aromatic
hydrocarbons, this division has been adopted in similar studies and is a step
towards analysing the underlying ISM chemistry \citep{Tielens1987, Dwek1998,
Weingartner2001}.

\subsubsection{Interstellar Dust Growth}\label{SEC:dust_growth}

The mass of dust in the ISM may increase over time, as gas-phase elements
collide with existing grains \citep{Draine1990}.  To model the accretion of
dust grains in the ISM, we follow the prescription of \citet{Dwek1998} and \citet{Hirashita1999} and in
every time step compute each cell's instantaneous dust growth rate
\begin{equation}
\left(\frac{\diff M_{i,\text{dust}}}{\diff t} \right)_\text{g} = \left( 1
- \frac{M_{i,\text{dust}}}{M_{i,\text{metal}}} \right) \left(
\frac{M_{i,\text{dust}}}{\tau_\text{g}} \right),
\label{EQN:growth_rate}
\end{equation}
where $M_{i,\text{dust}}$ is the cell's mass of species $i$ dust, summed over
components originating from AGB stars, SNe Ia, and SNe II, $M_{i,\text{metal}}$
is the corresponding species $i$ metal mass, and $\tau_\text{g}$ is a
characteristic growth timescale.  Note that the first factor in parentheses
induces a growth rate that depends on the local dust-to-metal ratio and slows
the accretion rate as gas-phase metals are condensed into dust \citep{Dwek1980,
McKee1989}.

Previous studies have explored accretion timescales dependent upon local gas
density and temperature \citep{Yozin2014, Zhukovska2014, Bekki2015a}
as well as metallicity \citep{Inoue2003a, Asano2013a} in an attempt to better
match the observed dust-metallicity relation.
Following these prescriptions, for each gas cell we compute the local
dust growth timescale
\begin{equation}
\tau_\text{g} = \tau_\text{g}^\text{ref} \left( \frac{\rho^\text{ref}}{\rho} \right) \left( \frac{T^\text{ref}}{T} \right)^{1/2},
\end{equation}
where $\rho$ and $T$ are the density and temperature of the gas cell,
$\rho^\text{ref}$ and $T^\text{ref}$ are reference values for density and
temperature in molecular clouds, and $\tau_\text{g}^\text{ref}$ is an overall
normalisation influenced by factors like atom-grain collision sticking
efficiency and grain cross section (see Section 2.2 of \citet{Hirashita2000}
for a detailed derivation).  We take $\rho^\text{ref}$ to be $1 \, \text{H
atom} \, \text{cm}^{-3}$ and $T^\text{ref} = 20 \, \text{K}$.  This growth
timescale is shortest in dense gas where collisions are more frequent than in
the diffuse halo. Future efforts will need to more accurately model atom-grain
collisions, taking into account a possibly temperature-dependent collision
sticking efficiency and a realistic grain size distribution
\citep{Weingartner2001, Li2001}.

\subsubsection{Grain Destruction}\label{SEC:dust_destruction}

Dust grains that have formed in the ISM can subsequently be destroyed through a
number of processes, including shocks from SN remnants \citep{Seab1983,
Seab1987, Jones1994}, thermal and nonthermal sputtering \citep{Draine1979b,
Tielens1994, Caselli1997}, and grain-grain collisions \citep{Draine1979a,
Jones1996}.  In particular, shocks reduce the net efficiency of dust formation
in SNe, expected to be the primary producers of dust at high redshift.  The
mass of grains destroyed in such a manner is thought to be proportional to the
energy of the shocks \citep{McKee1989}.

Paralleling Equation~(\ref{EQN:growth_rate}), for every active cell we can
estimate the local dust destruction rate for species $i$ as
\begin{equation}
\left( \frac{\diff M_{i,\text{dust}}}{\diff t} \right)_\text{d} = - \frac{M_{i,\text{dust}}}{\tau_\text{d}},
\label{EQN:destruction_rate}
\end{equation}
where $\tau_\text{d}$ is a characteristic destruction timescale
\citep{McKee1989, Draine1990, Jones2011} and again the dust mass is computed by
summing its components originating from AGB stars, SNe Ia, and SNe II.  While
we could tie $\tau_\text{d}$ to the galaxy-wide SN rate \citep{Hirashita2002}
or the radially-dependent gas surface density and SN rate \citep{Calura2008}, we
instead estimate a dust destruction timescale on a cell-by-cell basis, given by
\begin{equation}
\tau_\text{d} = \frac{M_\text{g}}{\epsilon \, \gamma \, M_\text{s}(100)},
\end{equation}
where $M_\text{g}$ is the gas mass within a cell, $\epsilon$ denotes
the efficiency with which grains are destroyed in SN shocks, $\gamma$ is
the local Type II SN rate, and $M_\text{s}(100)$ is the mass of gas
shocked to at least $100$ km s$^{-1}$ \citep{Dwek1980, Seab1983, McKee1989}.
We take $\epsilon = 0.3$, though there is a range of physically plausible
values used in previous work.  Applying the Sedov-Taylor solution to a
homogeneous environment yields
\begin{equation}
M_\text{s}(100) = 6800 \, E_\text{SNII,51} \left( \frac{v_\text{s}}{100 \; \text{km s$^{-1}$} }\right)^{-2} \; \text{M}_\odot,
\label{EQN:M_s}
\end{equation}
where $E_\text{SNII,51}$ is the energy released by a Type II SN in units
of $10^{51} \, \text{erg}$, and $v_\text{s} \sim 100$ km s$^{-1}$ is the shock
velocity \citep{McKee1989}.  Equation~\ref{EQN:M_s} implicitly
assumes SN shock expansion into a homogeneous medium of $n = 0.13 \,
\text{cm}^{-3}$, corresponding to the star formation density threshold in our
model.  We employ a fixed ISM density when calculating $M_\text{s}(100)$
because the detailed ISM multi-phase structure and, in particular, the ambient
gas density around each SN are not resolved in our simulations. In any
case, we caution that this destruction prescription neglects thermal
sputtering and grain-grain collisions in the CGM and thus may overdeplete
gas-phase metals in the diffuse halo.

During every time step, the net dust growth rate in every active gas cell
is computed by combining Equations~(\ref{EQN:growth_rate}) and
(\ref{EQN:destruction_rate}), and this rate is used to update the local dust
mass.  When performing this update, we keep the relative proportions of dust
mass originating from AGB stars, SNe Ia, and SNe II constant.

Additionally, the amount of dust in the ISM is reduced when star
particles are created.  We assume that dust and gas-phase metals are
distributed uniformly within each gas cell, so that each star particle's metals
come from gas-phase and dust sources in the same proportion as in the ISM.  For
example, if a star particle is created from a cell in which 25 per cent of oxygen
exists as dust and 75 per cent exists in the gas phase, then the amount of gas-phase
oxygen lost by the cell is three times greater than the amount of oxygen dust
lost.  Other schemes to consume dust during star formation may be possible but
more challenging to implement.

\subsubsection{Dust Effects on Cooling}

The galaxy formation model highlighted in Section~\ref{SEC:galaxy_formation}
calculates gas cooling rates using contributions from primordial species,
metals, and Compton cooling.  In particular, the metal-line cooling rate is
assumed to scale linearly with the gas-phase metallicity \citep[see Equation~(12)
in][]{Vogelsberger2013}.  The depletion of metals onto dust grains will
decrease the gas-phase metallicity of the ISM and therefore reduce metal-line
cooling.

Observations suggest that dust cooling channels can explain the formation of
very metal poor stars \citep{Caffau2011, Klessen2012}, and modelling finds
that cooling-induced fragmentation of dust impacts
low-metallicity star formation \citep{Schneider2006, Tsuribe2006, Dopcke2013}.
In numerical work, the temperature of dust grains can be computed via thermal
equilibrium calculations accounting for atomic line emission, grain
photoelectric effect heating, heating of dust through external radiation
fields, and dust cooling through thermal emission \citep{Krumholz2011}.  Dust
may affect cooling rates computed in cosmological simulations using local
ionising radiation \citep{Cantalupo2010, Gnedin2012, Kannan2014}.  In hot gas
with temperatures above $10^{6} \, \text{K}$, cooling from ion-grain collisions
is expected to be strong \citep{Ostriker1973}.  For simplicity of our model, we
neglect dust cooling channels in this first study.

While our model does not currently implement any specific dust cooling
processes, metal-line cooling will be reduced in comparison with previous
simulations involving \textsc{arepo}'s galaxy formation physics.  Decreased
cooling will in turn have a dynamical effect on galaxy formation, lowering the
star formation rate (SFR).  Future inclusion of dust cooling channels may
weaken some of the dynamical effects seen in this work.

\subsubsection{Winds from Stellar Feedback}

We employ the non-local stellar feedback implementation from
\citet{Vogelsberger2013} with some modification for dust.  In this feedback
prescription, gas cells in star-forming regions of the ISM are stochastically
converted into wind particles, given a velocity
\begin{equation}
v_\text{w} = \kappa_\text{w} \sigma_\text{DM}^\text{1D},
\label{EQN:v_wind}
\end{equation}
where $\kappa_\text{w}$ is a dimensionless scale and
$\sigma_\text{DM}^\text{1D}$ is the local one-dimensional dark matter velocity
dispersion, and allowed to move without hydrodynamic constraints.

When wind particles recouple to the gas, they deposit their metals, both
gas-phase and dust, in the same proportion as the ISM from which they were
launched.  Thus, these stellar winds help carry metals and drive outflows away
from dense regions of the ISM.  In Section~\ref{SEC:feedback}, we investigate
the importance of these winds and their strength in distributing dust
throughout a galaxy.

\section{Cosmological Runs}\label{SEC:cosmological_runs}

\subsection{Initial Conditions}

The initial conditions for our runs consist of the Aquarius suite of haloes,
previously used for high-resolution cosmological studies of Milky Way-sized
structure \citep{Springel2008}.  All runs adopt the $\Lambda$CDM cosmological
parameters $\Omega_\text{m} = \Omega_\text{dm} + \Omega_\text{b} = 0.25$,
$\Omega_\text{b} = 0.04$, $\Omega_\Lambda = 0.75$, $\sigma_8 = 0.9$, $n_s = 1$,
and Hubble constant $H_0 = 100 \, h \, \text{km} \, \text{s}^{-1} \,
\text{Mpc}^{-1} = 73 \, \text{km} \, \text{s}^{-1} \, \text{Mpc}^{-1}$, which
were employed in the Millennium and Millennium-II suites of simulations
\citep{Springel2005a, BoylanKolchin2009}.  Though recent results from the
Wilkinson Anisotropy Microwave Probe \citep{Hinshaw2013} and Planck satellite
\citep{PlanckCollaboration2014} suggest, for example, a value of $h < 0.73$, we
choose to adopt these parameters in part to remain consistent with recent
hydrodynamical simulations also utilising the Aquarius suite of haloes in
\textsc{arepo} \citep{Marinacci2014}.

The Aquarius haloes are labelled with the letters A through H, and each is
simulated in a periodic box of side length $100 \, h^{-1} \, \text{Mpc}$.  In
Aquarius nomenclature, our main simulations were performed using level 5
initial conditions.  In addition, we simulated the Aquarius C halo using higher
and lower resolution initial conditions, levels 4 and 6, respectively, in order
to study the convergence properties of the fiducial dust model.  We have found
our dust model to be well-converged and use the highest resolution initial
conditions for some visualisations in this work.  The Aquarius C halo, which we
also use to study variations in dust and feedback physics, was adopted in the
Aquila comparison project \citep{Scannapieco2012} to analyse the results of a
wide range of cosmological hydrodynamical codes.  This halo has a fairly
quiescent merger history, especially at low redshift \citep{Wang2011}.  We also
note that \citet{Marinacci2014} applied this galaxy formation model without
dust to the Aquarius haloes and found robust convergence over these same
resolution levels.

The same gravitational softening length was used for gas, dark matter, and
stars, and it was kept constant in comoving units until $z = 1$, with a maximum
value in physical units of $680 \, \text{pc}$ for the high-resolution region
using level 5 initial conditions.  The maximum values for level 4 and 6 initial
conditions were factors of two lower and higher, respectively.  The $z = 1$
softening length was then used down to $z = 0$.  We employ the \textsc{subfind}
algorithm \citep{Springel2001} for determining gravitationally-bound structure
and substructure.

\begin{table*}
 \centering
  \begin{tabular}{lrrrrrrrrrr}
   \hline
   Run & $R_{200}$ & $M_\text{gas}$ & $M_\text{dm}$ & $M_\text{dust}$ & $M_\text{gas,disc}$ & $M_\text{dust,disc}$ & $N_\text{gas}$ & $N_\text{dm}$ & $m_\text{gas}$ & $m_\text{dm}$ \\
   & $[\text{kpc}]$ & $[10^{10} \, \text{M}_\odot]$ & $[10^{10} \, \text{M}_\odot]$ & $[10^8 \, \text{M}_\odot]$ & $[10^{10} \, \text{M}_\odot]$ & $[10^8 \, \text{M}_\odot]$ & & & $[10^5 \, \text{M}_\odot]$ & $[10^5 \, \text{M}_\odot]$ \\
   \hline
   A5:FI & 237.5 & 11.62 & 182.29 & 5.61 & 4.52 & 3.79 & 211527 & 690478 & 5.03 & 26.40 \\
   B5:FI & 183.0 & 4.22 & 78.18 & 3.48 & 1.37 & 1.82 & 119386 & 518981 & 3.35 & 17.59 \\
   C5:FI & 233.5 & 11.11 & 175.93 & 5.98 & 3.47 & 3.47 & 253836 & 814834 & 4.11 & 21.59 \\
   D5:FI & 240.7 & 15.18 & 195.56 & 12.20 & 6.12 & 7.66 & 319074 & 846419 & 4.40 & 23.10 \\
   E5:FI & 206.7 & 6.14 & 114.16 & 5.35 & 1.09 & 1.00 & 179039 & 652270 & 3.33 & 17.50 \\
   F5:FI & 208.0 & 9.26 & 113.67 & 9.48 & 3.58 & 5.55 & 375068 & 942365 & 2.30 & 12.06 \\
   G5:FI & 201.9 & 9.94 & 95.99 & 7.78 & 5.20 & 5.74 & 317798 & 769854 & 2.83 & 14.88 \\
   H5:FI & 180.1 & 2.61 & 76.24 & 1.67 & 0.20 & 0.34 & 87369 & 588050 & 2.96 & 15.56 \\
   \hline
   C4:FI & 232.5 & 6.82 & 159.19 & 3.89 & 1.44 & 1.76 & 1265814 & 5898234 & 0.51 & 2.70 \\
   C6:FI & 235.6 & 13.92 & 179.84 & 6.91 & 4.97 & 4.54 & 39592 & 104118 & 32.90 & 172.73 \\
   \hline
  \end{tabular}
  \caption{Basic data on all eight Aquarius haloes simulated using level 5
  initial conditions and the fiducial dust model detailed in
  Section~\ref{SEC:fiducial}.  Additionally, we show the same statistics for
  the Aquarius C halo simulated at levels 4 and 6.  To keep consistent with the
  naming convention adopted later in this work, runs are referred to by their
  halo letter (A through H), resolution level (4, 5, or 6), and the suffix
  ``FI'' to denote that the fiducial dust physics were used.  Here, the virial
  radius $R_{200}$ is the radius about the halo's potential minimum enclosing a
  density 200 times greater than the critical density.  We list the gas, dark
  matter, and dust masses using all cells within the halo, denoted
  $M_\text{gas}$, $M_\text{dm}$, and $M_\text{dust}$, respectively.  Here,
  $M_\text{gas,disc}$ and $M_\text{dust,disc}$ are the
  estimated gas and dust masses in the galactic disc, computed
  using only dense ISM gas cells as determined by the temperature-density phase
  space cut in Equation~(\ref{EQN:Torrey2012}).  Finally, we indicate the number
  of gas and dark matter cells in the halo and their respective mass
  resolutions.}
  \label{TAB:ICs}
\end{table*}

In Table~\ref{TAB:ICs}, we provide basic statistics on these haloes, computed
at $z = 0$ using the fiducial model described below.  We provide two
computations of the gas and dust masses:
$M_\text{gas}$ and $M_\text{dust}$, which are summed over all
cells in the halo, and $M_\text{gas,disc}$ and $M_\text{dust,disc}$,
which include contributions only from dense ISM gas cells and
estimate the gas and dust content of the galactic disc.  We
isolate ISM gas by filtering cells according to temperature $T$ and density
$\rho$ using the relation
\begin{equation}
\log \left( \frac{T}{[\text{K}]} \right) < 6 + 0.25 \, \log \left( \frac{\rho}{10^{10} \, [\text{M}_\odot \, h^2 \, \text{kpc}^{-3}]} \right),
\label{EQN:Torrey2012}
\end{equation}
which has been shown to remove cells in the hot halo \citep[see Equation~(1)
in][]{Torrey2012}.  Additionally, Table~\ref{TAB:ICs} indicates the number and
mass resolution of gas and dark matter cells in each halo.

\subsection{Fiducial Parameters}\label{SEC:fiducial}

\begin{table*}
 \centering
  \begin{tabular}{lp{0.35\linewidth}p{0.4\linewidth}}
   \hline
   Variable & Fiducial Value & Description \\
   \hline
   $\delta^\text{AGB,$\text{C}/\text{O}>1$}_{i}$ & $0.0$ for $i = \text{H, He, N, O, Ne, Mg, Si, Fe}$, \newline $1.0$ for $i = \text{C}$ & dust condensation efficiency for species $i$ \newline in AGB stars with $\text{C}/\text{O} > 1$ in ejecta \\
   $\delta^\text{AGB,$\text{C}/\text{O}<1$}_{i}$ & $0.0$ for $i = \text{H, He, N, C, Ne}$, \newline $0.8$ for $i = \text{O, Mg, Si, Fe}$ & dust condensation efficiency for species $i$ \newline in AGB stars with $\text{C}/\text{O} < 1$ in ejecta \\
   $\delta^\text{SN}_{i}$ & $0.0$ for $i = \text{H, He, N, Ne}$, \newline $0.5$ for $i = \text{C}$, \newline $0.8$ for $i = \text{O, Mg, Si, Fe}$ & dust condensation efficiency for species $i$ in SNe \\
   $\tau_\text{g}^\text{ref}$ & 0.2 & reference dust growth timescale, in units of $[\text{Gyr}]$ \\
   $E_\text{SNII,51}$ & 1.09 & available energy per SN II, in units of $[10^{51} \, \text{erg}]$ \\
   \hline
  \end{tabular}
  \caption{Fiducial dust model parameters.  The dust condensation efficiencies
  regulate the fraction of metal ejecta for a chemical species that condenses
  into dust, as outlined in Section~\ref{SEC:dust_production}.  The
  reference growth timescale $\tau_\text{g}^\text{ref}$
  influences the rate at which gas-phase metals deplete onto dust grains, and
  $E_\text{SNII,51}$ controls the energy per SN II, which in turn affects grain
  destruction in SN shocks.}
  \label{TAB:fiducial}
\end{table*}

\begin{table*}
 \centering
  \begin{tabular}{rrrr}
  \hline
  $M \, [\text{M}_\odot]$ & $M_\text{dust}(Z = 0.004) \, [\text{M}_\odot]$ & $M_\text{dust}(Z = 0.008) \, [\text{M}_\odot]$ & $M_\text{dust}(Z = 0.02) \, [\text{M}_\odot]$ \\
  \hline
  2 & 0.065 & 0.025 & 0.009 \\
  4 & 0.003 & 0.006 & 0.015 \\
  6 & 0.006 & 0.011 & 0.029 \\
  12 & 0.392 & 0.381 & 0.365 \\
  20 & 0.780 & 0.781 & 0.780 \\
  \hline
  \end{tabular}
  \caption{Sample of condensed dust masses ($M_\text{dust}$) as a
  function of initial stellar mass ($M$) and metallicity ($Z$) using the
  stellar yields and dust condensation formulae in Section~\ref{SEC:methods}.
  We assume stars have scaled solar abundances.  In our simulations, the
  transition between AGB stars and SNe II occurs at $M = 8 \, \text{M}_\odot$.
  The yields for AGB stars are presented in \citet{Karakas2010} at $Z = 0.019$
  but listed here as $Z = 0.02$ for easy comparison with SNe yields at $Z =
  0.02$.}
  \label{TAB:masses}
\end{table*}

In Table~\ref{TAB:fiducial}, we present the set of parameter values that
comprise our fiducial dust model.  The fiducial feedback parameters are similar
to those used in \citet{Vogelsberger2013}, with $E_\text{SNII,51} = 1.09$
and $\kappa_\text{w} = 3.0$, and we did not retune the feedback model after
including dust.  Given that dust depletion will affect cooling rates and star
formation, future feedback modifications may be required.  This highlights the
need to include dust in detailed galaxy formation models.

The fiducial dust condensation efficiencies follow those from \citet{Dwek1998}
and assume that a certain fraction of ejecta from an AGB star or SN
exists as dust, with the remainder occupying the gas phase.  These efficiencies
are allowed to vary among species, since AGB stars and SNe are thought to
produce dust of differing compositions \citep{Ferrarotti2006, Zhukovska2008}.
For example, the high and low values of $\delta^\text{AGB,$\text{C}/\text{O}>1$}_\text{C}$ and
$\delta^\text{SN}_\text{C}$, respectively, are motivated by the idea that
carbon-rich AGB stars are the dominant producers of carbonaceous grains.
We caution that the condensation efficiencies used for SNe Ia are high
given recent observations and modelling work that suggest dust forms
inefficiently in SNe Ia \citep{Nozawa2011, Gomez2012}.  However, as shown in
Section~\ref{SEC:results}, SNe Ia produce only about 10 per cent of the dust in a
galaxy, and so our results are not very sensitive to the condensation
efficiencies for SNe Ia.

While condensation efficiencies of the form described in
Section~\ref{SEC:dust_production} have also been adopted in more recent
simulations \citep{Calura2008, Bekki2013}, stellar models
have begun to analyse the composition of ejected dust as a function of a star's
mass and metallicity \citep{Ferrarotti2006, Bianchi2007, Zhukovska2008,
Nanni2013, Schneider2014}.  For example, low-metallicity AGB stars are thought
to produce very low amounts of silicate grains \citep{Ferrarotti2006}.
Also, some of these recent models track dust grains for SiC,
Al$_2$O$_3$, or other compounds that do not map neatly onto our chemical
evolution model.  While we do not adopt condensation efficiencies that vary
with mass or metallicity, we can still compare the dust masses predicted by our
stellar yield tables and condensation efficiencies with those from more
detailed stellar models.  In Table~\ref{TAB:masses}, we calculate the total
dust masses condensed in our model for various choices of stellar mass and
metallicity.  We note that AGB stars produce total dust masses around $10^{-2}
- 10^{-3} \, \text{M}_\odot$ with no strong metallicity dependence, fairly
consistent with stellar models \citep{Zhukovska2008, Nanni2013}.

We can also compare the results of Table~\ref{TAB:masses} to predictions of
dust masses condensed in SNe II, although we caution that this subject is
debated in the literature.  For SNe II with masses of $12 \, \text{M}_\odot$
and $20 \, \text{M}_\odot$, we calculate condensed dust masses of roughly $0.4
\, \text{M}_\odot$ and $0.8 \, \text{M}_\odot$, respectively.  In comparison,
\citet{Todini2001} predicted about $0.3 \, \text{M}_\odot$ and $0.5 \,
\text{M}_\odot$ of dust for these choices of stellar mass.  Our results are
consistent with the estimate that between 2 per cent and 5 per cent of a progenitor's mass
becomes dust in SNe II in the range $13 \, \text{M}_\odot \leq M \leq 40 \,
\text{M}_\odot$ \citep{Nozawa2003}.  However, there is some tension with
observations: SN 2003gd, for example, was determined to have no more than $0.02
\, \text{M}_\odot$ of dust \citep{Sugerman2006, Meikle2007}, and other SNe have
failed to produce the amount of dust expected from models \citep{Rho2008,
Rho2009}.  As stellar models begin to implement more complex chemical reaction
networks \citep[e.g.][]{Cherchneff2010} and more observations are gathered,
better constraints can be placed on the dust masses condensed in core-collapse
SNe.

In the ISM, refractory grains are estimated to have destruction timescales of
$10^{8} - 10^{9} \, \text{yr}$, lower than the typical injection time of dust
from stellar sources \citep{Barlow1978, Draine1979a, Dwek1980, McKee1989}.
Mechanisms that allow metals to recondense in the ISM and counter the
destruction-injection imbalance are needed to regulate gas-phase abundances.
The growth timescale $\tau_\text{g}$ characterises dust growth in
molecular clouds and the associated probability that a gas-phase metal atom
will collide and stick to an existing grain \citep{Draine1990, Hirashita2000}.
Just as we tie the dust destruction timescale in each cell to its local SN
rate, we adopt a growth timescale dependent on local gas density and
temperature that enables dust to grow more quickly in dense ISM gas.
There is evidence that dust growth is particularly dominant over
stellar dust production in galaxies above a certain critical metallicity
\citep{Inoue2011, Asano2013a, Mancini2015}, and other models have begun to
explore the effect of density, temperature, and metallicity variations on local
growth timescales \citep{Inoue2003a, Bekki2015a}.  While models that
employ variable growth timescales sometimes assume a characteristic grain size,
grain density, and atom collision sticking efficiency, increasing the
number of estimated parameters, in this work we fixed such parameters
at typical Galactic values in a manner similar to Equation~(12) in
\citet{Hirashita2000} in order to capture the essential density and temperature
dependence in the growth timescale.  The resulting estimated reference
timescale $\tau_\text{g}^\text{ref}$ is shown in
Table~\ref{TAB:fiducial}.

\begin{table*}
 \centering
  \begin{tabular}{lll}
   \hline
   Name & Abbreviation & Physics \\
   \hline
   fiducial & FI & fiducial dust parameters from Table~\ref{TAB:fiducial}, $\kappa_\text{w} = 3.0$, includes AGN feedback \\
   \hline
   no feedback & NF & no stellar or AGN feedback \\
   AGN feedback & AF & only AGN feedback \\
   slow winds & SW & $\kappa_\text{w} = 1.85$ \\
   fast winds & FW & $\kappa_\text{w} = 7.4$ \\
   \hline
   constant growth timescale & CG & uses growth timescale of $\tau_\text{g} = 0.2 \, \text{Gyr}$, ignores local density and temperature \\
   constant destruction timescale & CD & uses destruction timescale of $\tau_\text{d} = 0.5 \, \text{Gyr}$, ignores local SN rate  \\
   low production & LP & all condensation efficiencies from Table~\ref{TAB:fiducial} halved \\
   no growth & NG & no growth mechanism \\
   no destruction & ND & no destruction mechanism \\
   production only & PO & no growth or destruction mechanisms \\
   \hline
   without dust & WD & no dust tracking, all metals exist in gas phase \\
   \hline
  \end{tabular}
  \caption{Description of different runs with varying feedback and dust model
  parameters.  The first and second columns characterise the nature of the
  model variation, and the third column specifies the exact physics changes.
  In particular, the no feedback (NF) and AGN feedback (AF) models ignore all
  stellar feedback and winds.  The slow winds (SW) and fast winds (FW) runs
  vary the dimensionless parameter $\kappa_\text{w}$ responsible for scaling
  the velocity of wind particles.  The constant destruction timescale (CD)
  variation adopts a constant value of $\tau_\text{d}$ used in
  Equation~(\ref{EQN:destruction_rate}), mimicking previous one-zone models that
  ignore local SNe when destroying grains.  The low production (LP) run
  adopts condensation efficiencies that are half of those of the fiducial model
  in an attempt to assess the importance of the grain formation process.  The
  without dust (WD) model uses fiducial feedback parameters but does not track
  dust.  It excludes all of the changes in Section~\ref{SEC:dust_model}.}
  \label{TAB:different_runs}
\end{table*}

Table~\ref{TAB:different_runs} lists the range of model variations that we
consider in addition to the fiducial setting.  These variations are divided
into two categories: those that apply the fiducial dust model and vary the
underlying feedback mechanisms, and those that adopt the fiducial feedback
settings and explore the impact of the various dust processes.  We also explore
the fiducial feedback model without dust tracking.  In this last model, all
metals are assumed to exist in the gas phase and contribute to metal-line
cooling.  This model without dust is expected to produce larger star formation
rates and gas-phase metallicities.

These feedback and dust model variations, together with the range of Aquarius
haloes used as initial conditions, provide a suitable starting point for
understanding the impact of dust on cosmological galaxy formation simulations.

In the remainder of this paper, we refer to our runs using four-character
identifiers: the first refers to the Aquarius halo chosen (i.e.\ one of A
through H), the second denotes the resolution level of the initial conditions
(i.e.\ 4, 5, or 6), and the last two indicate the underlying physical model
using the abbreviations in Table~\ref{TAB:different_runs}.  For example,
C5:ND is shorthand for the simulation of the Aquarius C halo using level 5
initial conditions with no dust destruction mechanism.

\section{Results}\label{SEC:results}

\subsection{Distribution of Dust in the Fiducial Model}

\begin{figure*}
\centering
\includegraphics{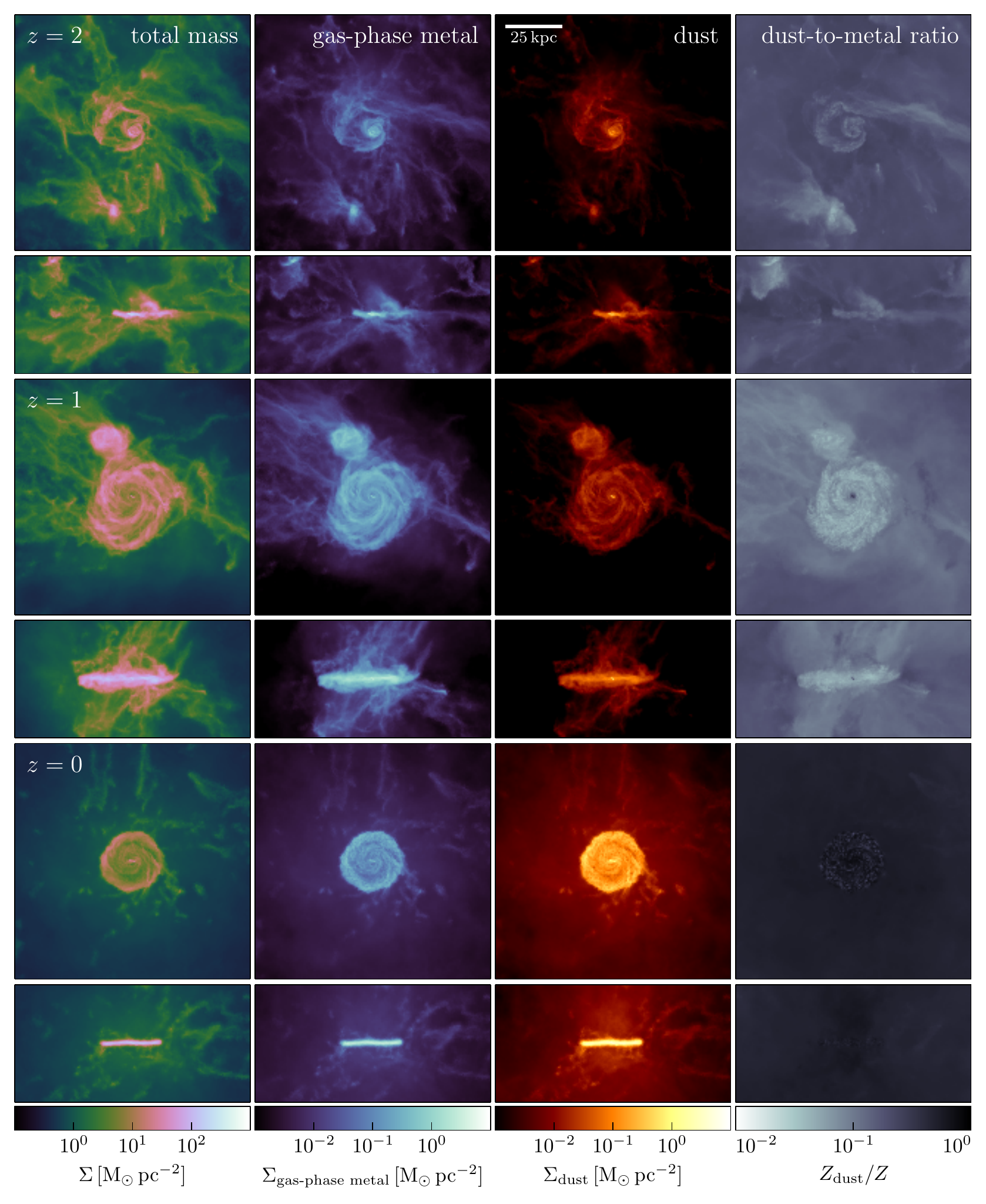}
\caption{Surface densities of baryons, gas-phase metals, and dust (first,
second, and third columns, respectively) as well as corresponding dust-to-metal
ratio (fourth column) for the Aquarius C halo at $z = 2$, $1$, and $0$ using
level 4 initial conditions with fiducial feedback and dust physics.  The two
images in each column at fixed redshift represent face-on (top) and edge-on
(bottom) projections.  All distances are given in physical units.  The scale
bar in the first row indicates $25 \, \text{kpc}$.  Projections were performed
in a cube of side length $150 \, h^{-1} \, \text{kpc}$ centered on the
potential minimum.  Movies of this simulation are available through
\protect\url{http://www.mit.edu/~ryanmck/\#research}.}
\label{FIG:projected_densities}
\end{figure*}

\begin{figure*}
\centering
\includegraphics{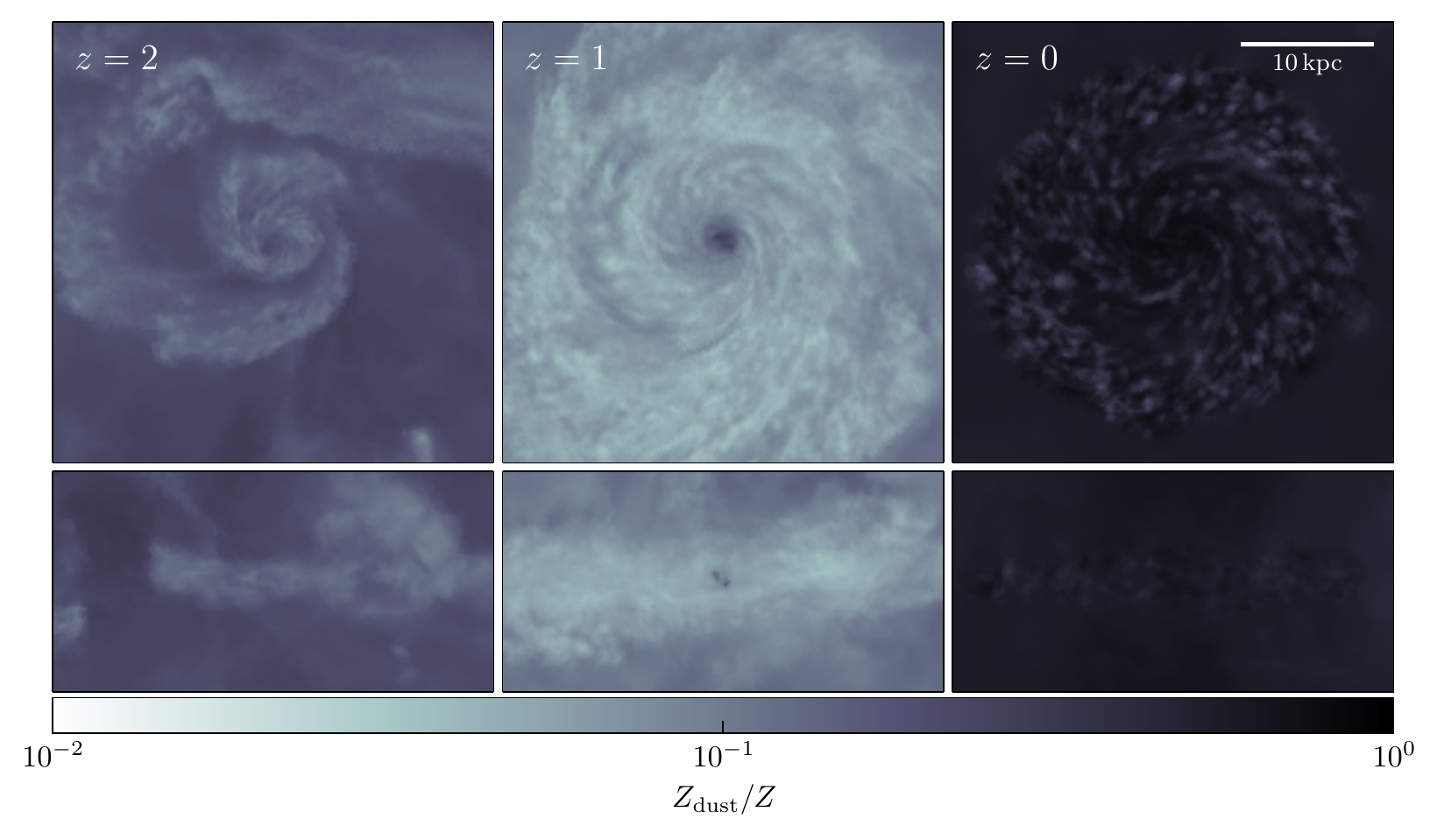}
\caption{Zoomed-in projections of the dust-to-metal ratio using the face-on
(top) and edge-on (bottom) views of the Aquarius C halo at $z = 2$, $1$, and
$0$ presented in Figure~\ref{FIG:projected_densities}.  These plots capture the
inner disc region, with the scale bar in the upper right indicating $10 \,
\text{kpc}$ in physical units.  A cube of side length $50 \, h^{-1} \,
\text{kpc}$ was used for the projection volume.}
\label{FIG:proj_Zdust_disk}
\end{figure*}

We first use our highest resolution simulation of the Aquarius C halo with
fiducial dust and feedback physics to analyse the distribution of gas-phase
metals and dust within the central halo and surrounding CGM.  This is motivated
by observations of significant amounts of dust in the CGM \citep{Menard2010,
Peeples2014, Peek2015} and the fact that gas-phase metals may condense into
dust in regions of low star formation.  In
Figure~\ref{FIG:projected_densities}, we show surface densities of baryons
($\Sigma$), gas-phase metals ($\Sigma_\text{gas-phase metal}$), and dust
($\Sigma_\text{dust}$) for face-on and edge-on views of the galactic disc at $z
= 2$, $1$, and $0$.  We also display the projected dust-to-metal ratio
($Z_\text{dust}/Z$) at these redshifts, where $Z_\text{dust}$ denotes the mass
fraction of dust and $Z$ is the usual metallicity, including gas-phase metals
and dust.  This halo does not undergo any major mergers below $z = 6$
\citep{Wang2011}, and so these images capture the quiescent formation of a disc
of diameter roughly $15 \, \text{kpc}$.

At $z = 2$, dust is most concentrated in the galactic center, with a surface
density roughly an order of magnitude larger than that outside the disc.  The
surface densities of gas-phase metals and dust evolve in a similar fashion from
$z = 2$ to $z = 1$ and largely trace the distribution of baryons.  By $z = 1$,
$\Sigma_\text{dust} > 10^{-2} \, \text{M}_\odot \, \text{pc}^{-2}$ extends
out to around $25 \, \text{kpc}$ as dust grows on the edges
of the galactic disc.  The surface density of gas-phase metals in the CGM
is largely unchanged from $z = 2$ to $z = 1$, with a
dust-to-metal ratio below $0.1$.  The dust-to-metal ratio is largest
near the galactic center and several times higher than the typical
value away from the disc.  By $z = 0$, there is less variation between
gas-phase metals and dust.  The central disc, roughly
$15 \, \text{kpc}$ across, has $\Sigma_\text{dust} > 10^{-1} \,
\text{M}_\odot \, \text{pc}^{-2}$, and the dust-to-metal ratio remains
highest in the star-forming central core where dust is
produced.  We caution that the absence of thermal sputtering in the
diffuse halo may artificially cause dust-to-metal ratios to rise in the CGM.

These results indicate that dust depletion can affect the distribution of
gas-phase metals, especially in regions where high star formation
activity produces the most dust.  To highlight the dust evolution in
the inner galactic disc, in Figure~\ref{FIG:proj_Zdust_disk} we display
zoomed-in projections of the dust-to-metal ratio for the face-on images from
Figure~\ref{FIG:projected_densities}.  At each of $z = 2$, $1$, and $0$, the
dust-to-metal ratio is largest near the galactic center.  However, by
$z = 0$, the spatial variation in dust-to-metal ratio that is prominent
at $z = 1$ diminishes, although there are small pockets of low dust-to-metal
ratio in the outer disc.  In conjunction with
Figure~\ref{FIG:projected_densities}, this suggests that observations
of galaxies should see dust-to-metal ratios that are highest
near galactic cores and vary with redshift.

\subsection{Impact of Feedback}\label{SEC:feedback}

\begin{figure*}
\centering
\includegraphics{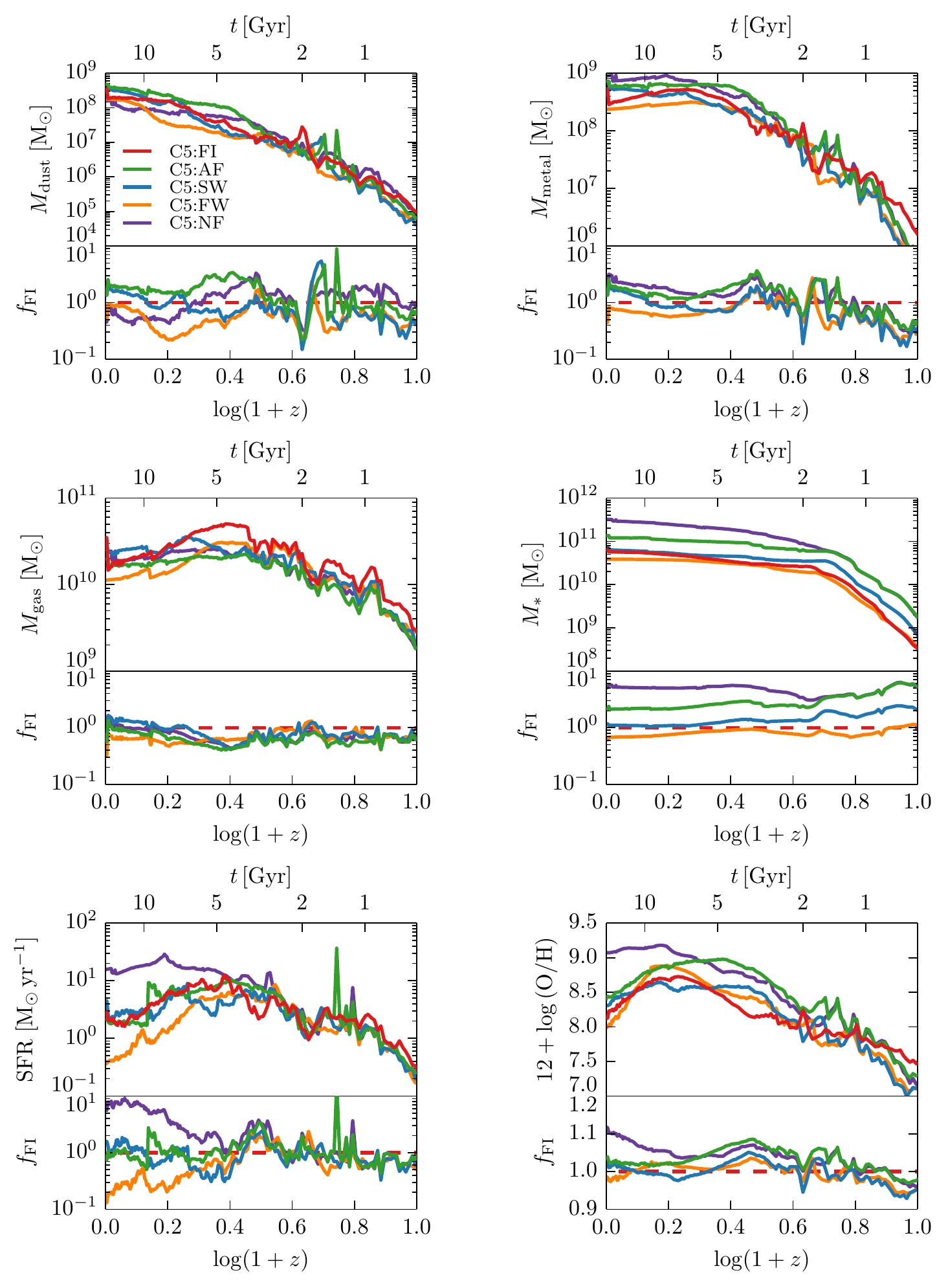}
\caption{Clockwise from top left, a comparison of dust mass, metal mass,
stellar mass, gas-phase metallicity, star formation rate, and gas mass as
a function of redshift for the Aquarius C galaxy using various feedback
parameters.  These quantities include contributions only from the dense
galactic disc and not the hot halo.  In each plot, the top panel shows results
on an absolute scale, while the ratio of quantities relative to the fiducial
run (denoted $f_\text{FI}$) is shown in the bottom panel.  This convention is
also adopted in subsequent figures.  The metal mass includes both gas-phase and
dust contributions.  The fast winds (FW) and no feedback (NF) runs produce dust
masses at $z = 0$ roughly half that produced in the fiducial (FI) model.  The
fast winds run sees the strongest decline in SFR and gas-phase metallicity
towards $z = 0$.}
\label{FIG:dust_quantities_comparison_feedback}
\end{figure*}

\begin{figure*}
\centering
\includegraphics{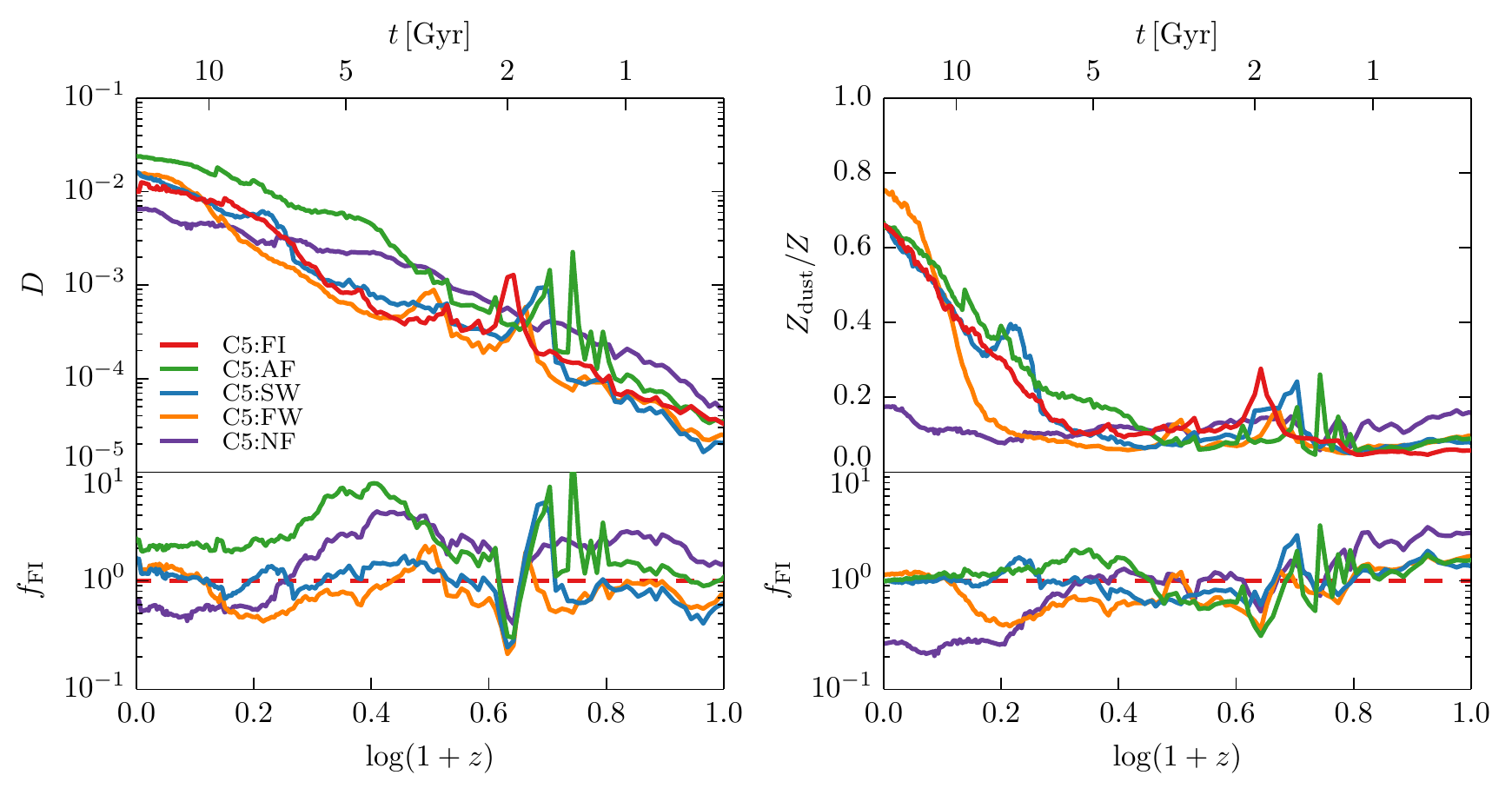}
\caption{Comparison of dust-to-gas ratio ($D$; left) and dust-to-metal ratio
($Z_\text{dust}/Z$; right) for the range of feedback models run on the Aquarius
C halo.  Quantities correspond to the dense galactic disc, as in
Figure~\ref{FIG:dust_quantities_comparison_feedback}.  The fiducial (FI), slow
winds (SW), and fast winds (FW) runs with stellar feedback yielded similar
dust-to-gas ratios at all redshifts.  There is significant scatter in the
dust-to-metal ratio, though the feedback run with the largest star formation
rate and gas-phase metallicity at $z = 0$ displayed the smallest dust-to-metal
ratio.  This is consistent with our SN-driven dust destruction
mechanism.}
\label{FIG:dust_ratio_comparison_feedback}
\end{figure*}

While some previous cosmological simulations treating dust have investigated
the effect of galactic winds \citep{Zu2011} and SN feedback
\citep{Bekki2015a}, there has been no comprehensive analysis of feedback
physics on dust evolution.  In
Figure~\ref{FIG:dust_quantities_comparison_feedback}, we show the total dust,
metal, gas, and stellar masses as a function of redshift for each of the
feedback variations detailed in Table~\ref{TAB:different_runs} applied to the
Aquarius C galaxy.  We also plot the SFR and gas-phase metallicity,
two dynamical quantities that will be impacted by the presence of dust.

Until $z \approx 4$, the no feedback and AGN-only feedback models are very
similar, producing stellar masses several times larger than those seen in the
fiducial, slow winds, and fast winds models with stellar feedback enabled.
This is consistent with the SFR plot, which shows that the no feedback
and AGN-only feedback runs have similar behaviour at high redshift.
All runs yielded dust masses at high redshift more than an
order of magnitude smaller than the several $10^{7} \, \text{M}_\odot$ and
above of dust observed for SDSS J1148+5251 and A1689-zD1 at $z = 6.4$ and
higher \citep{Valiante2009, Valiante2011, Watson2015}.  This is not too
surprising since simulations of haloes larger than those in the Aquarius suite
seem to be required to produce such dusty systems at high redshift.

\begin{figure}
\centering
\includegraphics{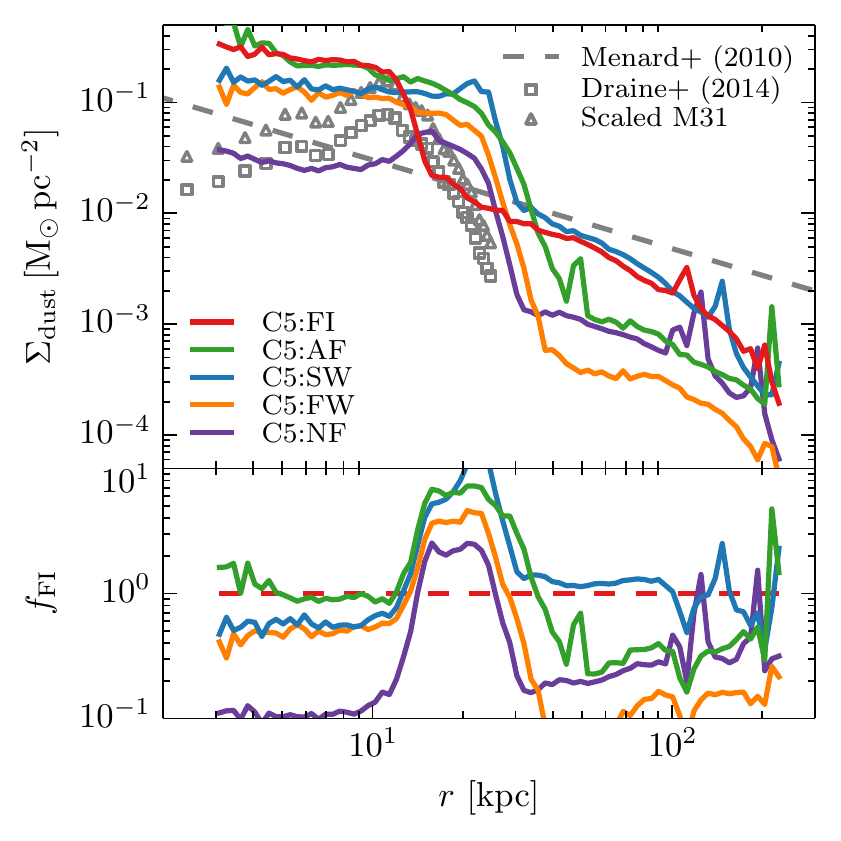}
\caption{Surface density of dust ($\Sigma_\text{dust}$) as a function of radial
distance from the Aquarius C halo's spin axis at $z = 0$ for our set of
feedback variations.  The radial range extends to $R_{200}$, as defined in
Table~\ref{TAB:ICs}, and the bottom subpanel shows surface density relative to
the fiducial result.  The gray dashed line indicates the $\Sigma_\text{dust}
\propto r^{-0.8}$ scaling observed in SDSS data over cosmological distances
\citep{Menard2010}.  The amplitude of this scaling has been chosen to align
with simulated data from $25 \, \text{kpc}$ out to roughly $100 \, \text{kpc}$.
Squares indicate measurements of M31's projected dust profile
\citep{Draine2014}.  Triangles denote the M31 data scaled by a factor of two to
model a Milky Way dust mass that may be higher than that of M31.}
\label{FIG:dust_radial_density_feedback}
\end{figure}

\begin{figure}
\centering
\includegraphics{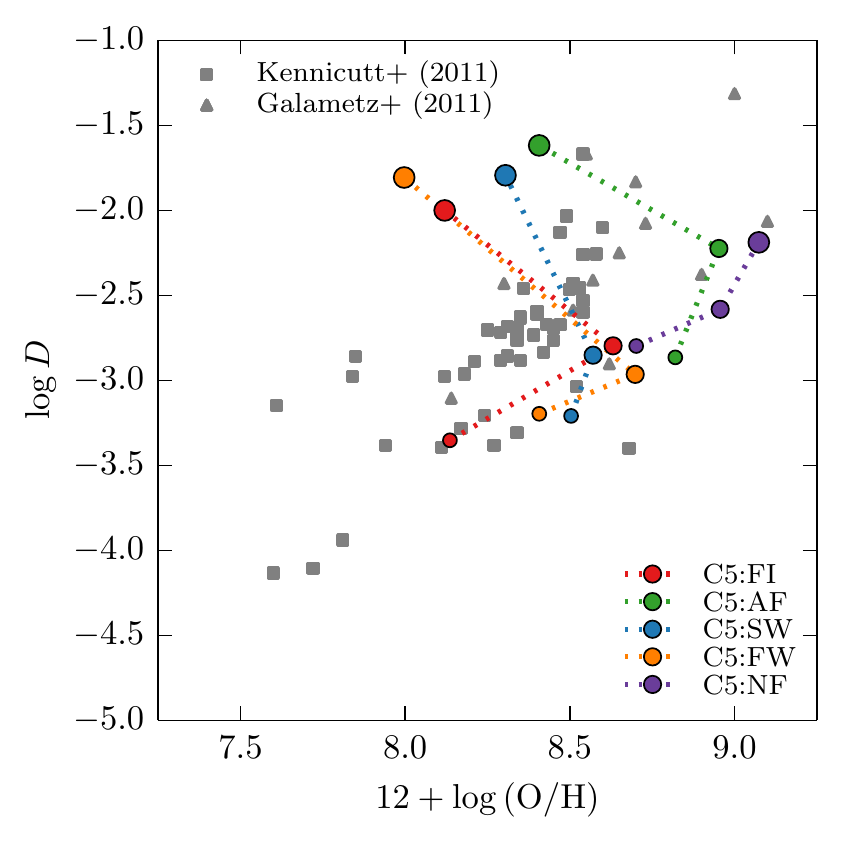}
\caption{Dust-to-gas ratio ($D$) versus gas-phase metallicity for varying
feedback runs involving the Aquarius C halo, plotted at $z = 2, 1,$ and $0$.
Smaller circles denote higher redshift.  The dust-to-gas ratio and gas-phase
metallicity have been computed using only dense gas cells in the galactic disc.
For comparison, observational data from \citet{Kennicutt2011} and
\citet{Galametz2011} \citep[as compiled in][]{RemyRuyer2014} are provided.  All
runs except that with no feedback experience a drop in gas-phase metallicity
from $z = 1$ to $z = 0$.}
\label{FIG:dust_metallicity_feedback}
\end{figure}

\begin{figure}
\centering
\includegraphics{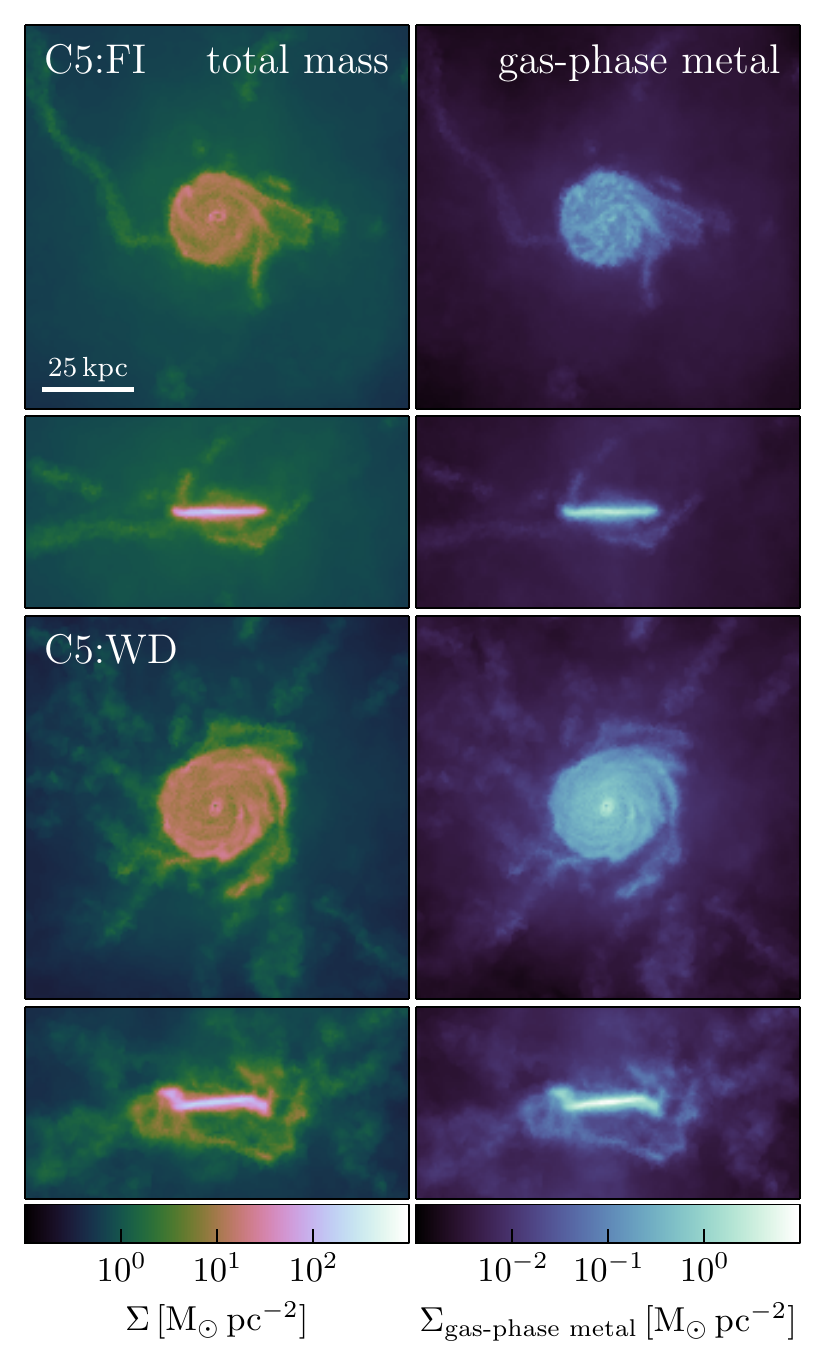}
\caption{Face-on and edge-on projections of total surface density (left column)
and gas-phase metal surface density (right column) at $z = 0$ for simulations
of the Aquarius C halo with fiducial dust (FI; top half) and without dust (WD;
bottom half).  The scale bar in the first row marks $25 \, \text{kpc}$, and the
projection volume is the same as that used in
Figure~\ref{FIG:projected_densities}.  Not shown are metals occupied by dust,
which exist only in the fiducial simulation.  Compared to the simulation
without dust, the fiducial dust run results in a smaller disc and a decreased
gas-phase metal surface density beyond $15 \, \text{kpc}$ from the galactic
center.}
\label{FIG:proj_withoutdust}
\end{figure}

At low redshift, AGN feedback strongly suppresses the SFR in the Aquarius C
galaxy.  The reduction in SFR provides more opportunity for gas-phase metals in
the ISM to condense into dust, leading to a decline in gas-phase metallicity
relative to the no feedback model.  The runs with stellar feedback see a
similar suppression of gas-phase metallicity below $z \approx 1$.  This effect
is most pronounced for the fast winds model, which is most efficient at driving
gas-phase metals away from star-forming regions and has the lowest stellar
mass.  In the absence of SN activity, these gas-phase metals can accrete onto
dust grains more rapidly.  The fast winds run sees a drop in gas-phase
metallicity of roughly $0.7 \, \text{dex}$ from $z = 1$ to $z = 0$.
We note that the gas-phase metallicities shown in
Figure~\ref{FIG:dust_quantities_comparison_feedback} correspond to the
dense galactic disc and that the hot halo sees less metal enrichment.  The
fiducial and slow winds models do not experience such severe declines
in gas-phase metallicity and are more similar to the AGN-only feedback run at
$z = 0$ than the extreme no feedback model.

Figure~\ref{FIG:dust_ratio_comparison_feedback} shows the evolution of the
dust-to-gas ratio $D$ and dust-to-metal ratio $Z_\text{dust}/Z$ for the
same feedback variations.  The fiducial model for this Milky Way-sized galaxy
yields $D \approx 10^{-2}$, consistent with estimates of the Galactic
value \citep{Gilmore1989, Sodroski1997, Zubko2004}.  The feedback variations
also yield dust-to-gas ratios within a factor of several of the fiducial value.
The dust-to-gas ratio increases by roughly $1 \, \text{dex}$ from $z = 2$ to
$z = 0$ for the fiducial, slow winds, and fast winds runs.
Above, we noted that the fast winds model promotes gas-phase depletion of
metals, and so one might expect a high dust-to-gas ratio.  However, because
star formation is so strongly suppressed in this model, the total amount of
dust is roughly half that of the fiducial run.  As a result,
the fast winds model has a dust-to-gas ratio comparable to the fiducial value.
On the other hand, the fast winds $z = 0$ dust-to-metal ratio is the largest of
all models, at nearly $0.75$.  This is slightly below the fiducial
model's $z = 0$ dust-to-metal ratio of near $0.65$, which is still
above but closer to estimates of the Galactic value.  The no
feedback run with more star formation was more effective at
regulating the dust-to-metal ratio and returning dust to the gas phase.

The dust-to-metal ratio is expected to increase significantly at low redshift:
previous modelling predicts $Z_\text{dust}/Z$ at $z = 0.5$, $1$, and $2$ to be
$50$, $30$, and $20$ per cent of the $z = 0$ value \citep{Inoue2003a}.  While
the feedback variations do not reproduce these precise numbers, there is a
noticeable increase towards low redshift.  This increase is present across all
feedback variations, even though the $z = 0$ dust-to-metal ratios range from
$0.2$ for the no feedback model to over $0.7$ for the fast winds model.  The
presence of feedback affects the normalisation of the dust-to-metal ratio but
does not strongly alter the shape of its evolution.

While the dust-to-metal ratio was strongly affected by variations in feedback,
other quantities, like surface density of dust projected onto the galactic
plane, are less sensitive.  In Figure~\ref{FIG:dust_radial_density_feedback},
we show the dust surface density $\Sigma_\text{dust}$ versus radial distance at
$z = 0$ for each feedback model.  Outside of the galactic disc, we
compare with the $\Sigma_\text{dust} \propto r^{-0.8}$ scaling observed in SDSS
data \citep{Menard2010}.  Additionally, since the Aquarius suite of haloes forms
Milky Way-sized galaxies, we overlay recent observations of the dust surface
density in the inner disc of M31, which has an estimated dust mass of $5.4
\times 10^{7} \, \text{M}_\odot$ \citep{Draine2014}.  In contrast to previous
figures, all feedback variations yield similar results, with the surface
density profiles peaking around $\Sigma_\text{dust} \sim 10^{-1} \,
\text{M}_\odot \, \text{pc}^{-2}$ for $z = 0$.  These surface densities display
moderate increases from $z = 1$ to $z = 0$, consistent with the growth in dust
mass from Figure~\ref{FIG:dust_quantities_comparison_feedback}.  All feedback
runs experience a drop in dust surface density of roughly one order of
magnitude or more when leaving the galactic disc, with normalisations and
shapes in broad agreement with M31 data.  These models also all predict
$\Sigma_\text{dust}$ scalings out to more than $100 \, \text{kpc}$ that are
fairly consistent with SDSS observations.  However, cosmological simulations of
larger galaxy populations will be needed to confirm this trend on the scales
detailed in \citet{Menard2010}.  Such simulations will also be able to
investigate predictions about the dust content of the intergalactic medium
\citep{Aguirre1999, Inoue2003b, Petric2006, Dijsktra2009, Corrales2012}.

Figure~\ref{FIG:dust_metallicity_feedback} displays the dust-metallicity
relation for these feedback runs, plotting dust-to-gas ratio versus gas-phase
metallicity at $z = 2$, $1$, and $0$.  We also display observations from
\citet{Kennicutt2011} and \citet{Galametz2011}, which cover roughly a $1.5 \,
\text{dex}$ range in oxygen abundance.  To facilitate comparison with these
data, we compute the dust-to-gas ratio and gas-phase metallicity using only
dense gas in the galactic disc, as determined by the temperature-density phase
space cut in Equation~(\ref{EQN:Torrey2012}).  We use this temperature-density
cut in all subsequent plots involving comparison to observational data.

Variations in feedback affect both the gas-phase metallicity and the
dust-to-gas ratio, particularly at low redshift.  All models except the no
feedback model see a decline in gas-phase metallicity from $z = 1$ to $z = 0$,
with the effect most pronounced for the fast winds run.  The fiducial
model is most similar to the runs with fast and slow winds, consistent with the results
in previous figures.  Although these stellar feedback models lie above
the dust-metallicity relation at $z = 0$, the fits at $z = 1$ and $z = 2$ are
much more reasonable.  The no feedback model has a high gas-phase metallicity
at all redshifts and thus lies slightly below the dust-metallicity
observations.  These results indicate that the inclusion of
feedback in our dust model can affect the dust-metallicity evolution of
galaxies and help limit high-redshift metallicities.

\begin{figure}
\centering
\includegraphics{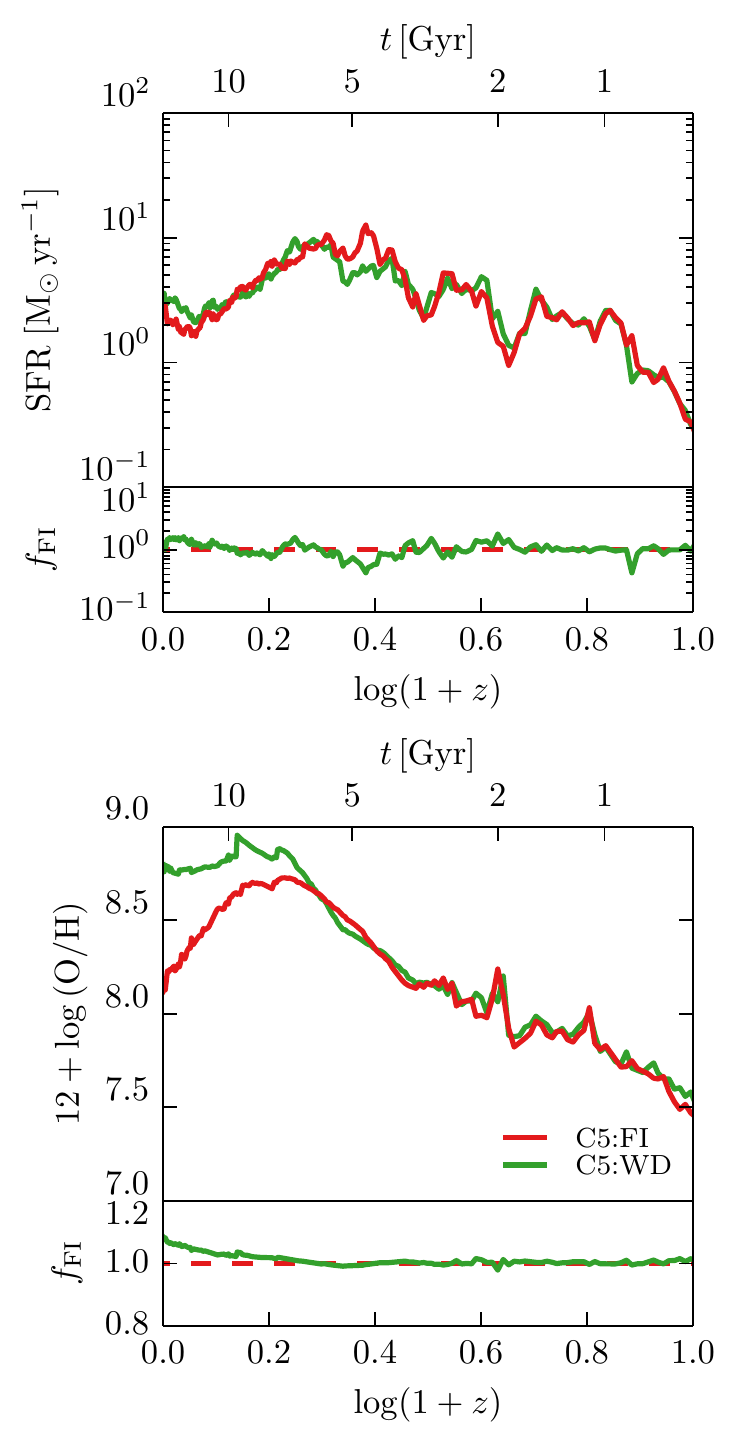}
\caption{A comparison of star formation rate (SFR; top) and gas-phase
metallicity (bottom) computed for the Aquarius C galaxy as a function of
redshift in the fiducial (FI) and without dust (WD) models.  The FI run has
decreased metal-line cooling, as gas-phase metals deplete onto dust grains.
The SFRs largely agree, but at $z = 0$, the difference in gas-phase
metallicities between the FI and WD runs is roughly $0.5$ dex.}
\label{FIG:dust_comparison_feedback_nodust}
\end{figure}

Because heavy elements trapped in dust grains do not contribute to metal-line
cooling in our model, we expect star formation to decrease as gas-phase metals
accrete onto dust.  This qualitative effect is shown in
Figure~\ref{FIG:proj_withoutdust}, which compares the projected baryon and
gas-phase metal densities for our fiducial model as well as a model with all
dust physics disabled.  In this latter model, all metals outside of stars are
considered to be in the gas phase.  The run without dust produces a
slightly larger galactic disc at $z = 0$, while the run with dust
slightly reduces the density of gas-phase metals in the CGM.  While we caution
that our fiducial dust model ignores dust sputtering and may overdeplete
gas-phase metals in the halo, these results suggest that the presence of dust
affects how galaxies evolve in simulations, even if feedback settings are
unaltered.

To analyse this effect quantitatively, in
Figure~\ref{FIG:dust_comparison_feedback_nodust} we plot the SFR and gas-phase
metallicity versus redshift for our fiducial model as well as our model without
dust.  Differences in these models are most noticeable beginning at $z \approx
1$.  While both models produce similar SFRs, the deviation in gas-phase
metallicity from $z \approx 1$ to $z = 0$ is more pronounced.  Without dust,
the gas-phase metallicity increases and plateaus, whereas the model with dust
sees a decline of roughly $0.5 \, \text{dex}$.  At low redshift, even the
fiducial feedback settings are sensitive to whether dust is included or not.

In the analysis above, the fiducial feedback model produced dust-to-gas and
dust-to-metal ratios in rough agreement with the Galactic values.
While its location on the observed dust-metallicity relation was in
tension with observations at $z = 0$, the fiducial run at higher redshift
yielded more accurate results.  The no feedback model, already known to
overproduce high stellar mass galaxies \citep{Vogelsberger2013}, has the
highest SFR and lowest dust-to-metal ratio of the various models.  Fast winds
strongly suppress star formation and dust production, carrying metals to
galactic regions where dust destruction is less effective.  This resulted in a
poor $z = 0$ dust-metallicity fit, owing to low redshift gas-phase depletion.
While the models with slow winds and AGN-only feedback do not suffer such
extreme problems, they offer no advantage over the fiducial feedback model.
Previous studies have shown that in the absence of winds, the dust-to-metal
ratio required to match observed intergalactic reddening is unphysical
\citep{Zu2011}.  We note that variations in stellar wind feedback affect our
dust-metallicity relation more strongly than in \citet{Bekki2015a}, though in
both cases stronger feedback suppresses the dust-to-gas ratio and gas-phase
metallicity.  In the remainder of this paper, we adopt the fiducial feedback
parameters.

\subsection{Variations of the Dust Model}\label{SEC:variations_dust}

\begin{figure}
\centering
\includegraphics{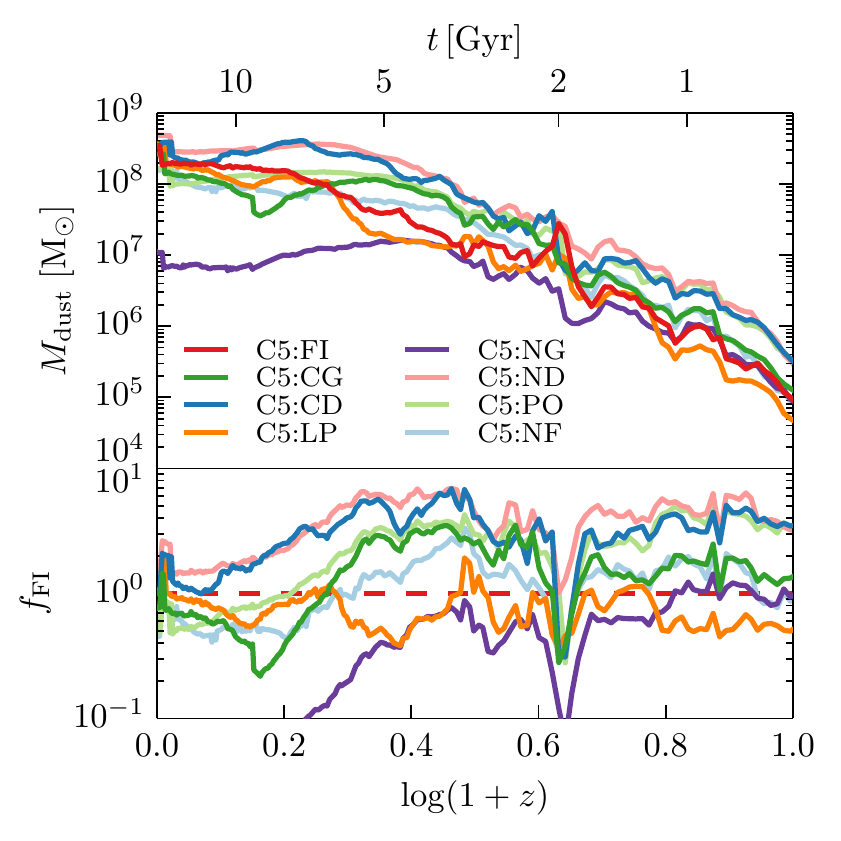}
\caption{Dust mass as a function of redshift for a variety of dust models
applied to the Aquarius C galaxy.  This only includes contributions from the
inner disc and not the hot hao.  The no feedback (NF) run is shown for
comparison.  At $z = 0$, the dust model without a growth mechanism (NG) yields
a final dust mass more than an order of magnitude below those obtained by
models with growth enabled.  The difference between a destruction mechanism
based on local SNe (FI) and a uniform timescale (CD) is minor at $z = 0$, but
at high redshift the CD run is more similar to the run without any destruction
(ND).  The model without any growth or destruction (PO) sees a dust mass at low
redshift reduced by roughly a factor of two compared to the ND run with growth.
The fiducial model was most similar to the low production (LP) model with
smaller condensation efficiencies.}
\label{FIG:dust_tot_comparison_Ccomp}
\end{figure}

Motivated by observations of high-redshift dusty galaxies, previous one-zone
dust models have studied the importance of dust growth in the ISM relative to
contributions from stellar sources.  Dust growth appears to be dominant in
galaxies above a critical metallicity, and galaxies with a shorter star
formation timescale reach this critical metallicity more quickly
\citep{Inoue2011, Asano2013a}.  The carbon-to-silicate dust mass ratio changes
significantly as dust growth overtakes stellar injection, and dust growth in
the ISM can be important even in dwarf galaxies \citep{Zhukovska2014}.  While
we do not investigate it in this paper, we acknowledge that variations in the
IMF can affect dust model results, particularly in low mass galaxies
\citep{Gall2011a}.  Top-heavy IMFs may also strengthen dust production from
SNe, despite leading to increased grain destruction \citep{Dunne2011}.  It is
therefore natural to consider how the various components of our dust model
impact galaxy evolution.  This will help to determine which aspects of dust
physics are most important and worthy of more detailed modelling in future
work.

In Figure~\ref{FIG:dust_tot_comparison_Ccomp}, we plot the redshift evolution
of dust mass in the Aquarius C galaxy for each of the dust model
variations listed in Table~\ref{TAB:different_runs}.  We also show how each
variation compares to the fiducial result.  Because dust was shown to have a
dynamical effect on star formation in Section~\ref{SEC:feedback}, the total
metal masses for each run show minor differences.  The dust mass evolution of
the fiducial model was most similar to that of the model with decreased dust
condensation efficiencies, enabling more mass to be ejected in the gas phase.
In fact, their $z = 0$ dust masses were almost identical, while at $z
\approx 2.5$ the fiducial model produced about twice as much dust.  This perhaps
reduces the effect of large uncertainties in dust condensation efficiencies,
especially for results at low redshift.  For modelling at high redshift,
metallicity-dependent condensation efficiencies \citep[as in the one-zone model
of][]{Zhukovska2008} may have more of an impact.

\begin{figure*}
\centering
\includegraphics{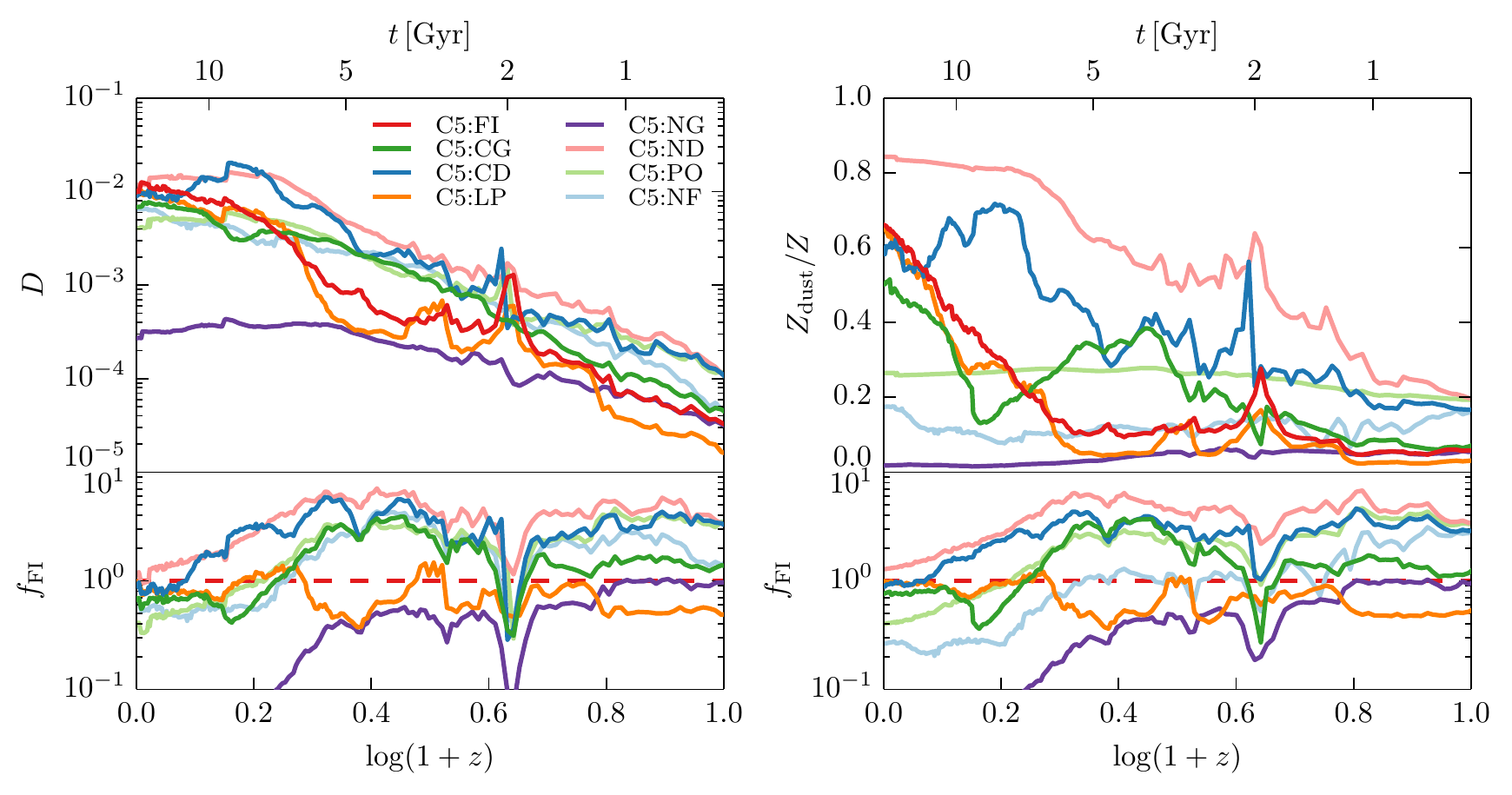}
\caption{Comparison of dust-to-gas ratio ($D$; left) and dust-to-metal
ratio ($Z_\text{dust}/Z$; right) for the range of dust models in
Figure~\ref{FIG:dust_tot_comparison_Ccomp} run on the Aquarius C galaxy.  These
ratios are computed using total masses within the dense galactic disc.  The no
feedback (NF) model is also shown for comparison.  In both plots, the model
lacking a dust destruction mechanism (ND) yielded the highest dust-to-gas and
dust-to-metal ratios at essentially all redshifts.  Conversely, the model with only dust
production enabled (PO) results in dust-to-gas and dust-to-metal ratios at $z =
0$ roughly half those obtained by the fiducial model.  The run without dust
growth but with destruction (NG) more strongly suppresses both ratios.  The NG
and PO runs display very flat dust-to-metal ratio evolutions.}
\label{FIG:dust_ratio_comparison_Ccomp}
\end{figure*}

In contrast, the model without dust growth deviated significantly from the
fiducial run, and the dust mass increased by only a factor of four from $z
\approx 4$ to $z = 0$.  The fiducial model and other models with growth enabled
saw an increase in dust mass closer to one order of magnitude.
This corroborates findings from earlier one-zone models and highlights the
importance of dust growth at low redshift.  We caution that more detailed
growth mechanisms following one-zone models that take into account variations
in dust grain sizes \citep{Hirashita2011, Asano2013b, Nozawa2015} and their
impact on grain collision rates will offer a better assessment of the role
dust growth plays in the ISM.  Before $z \approx 5$, the difference between a
model employing only stellar dust production and a model with both stellar
production and ISM dust growth is minor, while both models yield more dust than
in the fiducial run with destruction.  However, by $z = 0$, the model lacking
growth and destruction sees a final dust mass approximately two times smaller
than that obtained by the model with growth but without destruction.  Grain
growth becomes increasingly important towards low redshift.  This is consistent
with the hypothesis that galaxies pass through a critical metallicity at which
dust growth in the ISM dominates stellar production.

The run adopting a constant destruction timescale of $\tau_\text{d} = 0.5 \,
\text{Gyr}$ for dust destruction, rather than the fiducial timescale computed
locally using SN activity, behaves more like the model lacking any
destruction whatsoever and overproduces dust at high redshift relative to the
fiducial model.  By $z = 0$, the constant destruction timescale run agrees well
with the fiducial run.  Dust models that globally adopt $\tau_\text{d}
= 0.5 \, \text{Gyr}$ \citep[e.g.~the fiducial runs in][]{Bekki2015a} may be
overestimating a galaxy's dust content by a factor of several at high redshift.

\begin{figure*}
\centering
\includegraphics{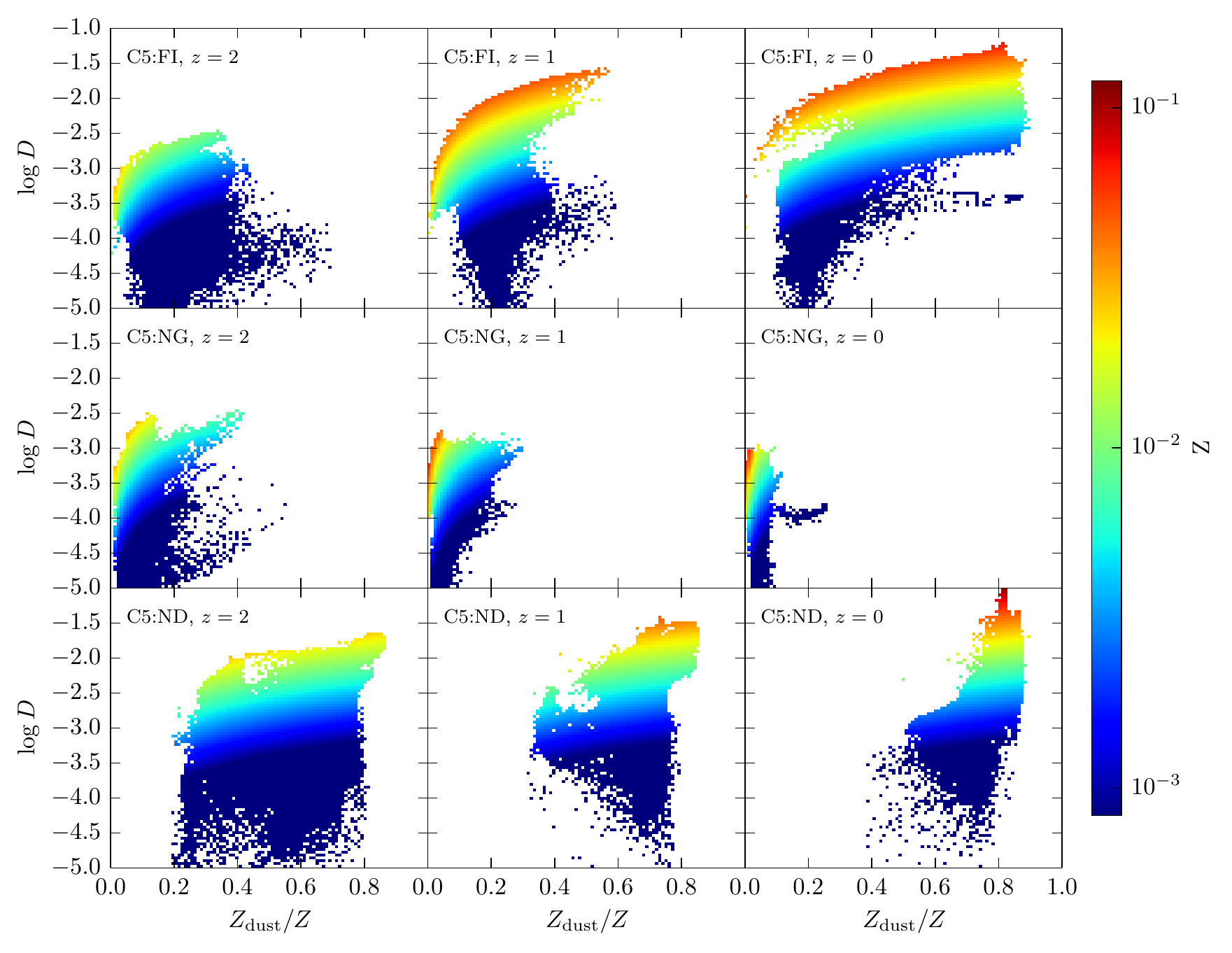}
\caption{Two-dimensional histogram of dust-to-gas ratio ($D$) versus
dust-to-metal ratio ($Z_\text{dust}/Z$) for all gas cells in the Aquarius C
halo at $z = 2$, $1$, and $0$ (left, middle, and right columns, respectively)
under the fiducial (FI; top row), no growth (NG; middle), and no destruction
(ND; bottom) models.  Bins are colored with the mass-weighted metallicity of
constituent gas cells.  In all cases, the highest metallicity regions tend to
have high dust-to-gas ratio, but the absence of a growth mechanism for the NG
model strongly suppresses an increase in the dust-to-metal ratio at all
redshifts relative to the fiducial model.  The presence of grain destruction in
the NG model yields a peak dust-to-gas ratio more than an order of magnitude
below that found in the ND model lacking a destruction mechanism.}
\label{FIG:dust_to_gas_metal_Ccomp}
\end{figure*}

Figure~\ref{FIG:dust_ratio_comparison_Ccomp} shows the dust-to-gas and
dust-to-metal ratios for these dust model runs, using contributions from
cells within the disc of the Aquarius C galaxy.  Nearly all
dust variations yield a $z = 0$ dust-to-gas ratio within a factor of two of the
fiducial value of $D \approx 10^{-2}$.  At high redshift, models
without dust destruction display an increased dust-to-gas ratio.  A dust model
without growth results in a late time dust-to-gas ratio suppressed by over $1
\, \text{dex}$ relative to the fiducial run.  Without growth, the dust-to-gas
ratio shows little evolution from its value at $z \approx 5$.  The effect of
limiting dust growth is much more severe than that of limiting grain
destruction.  Modelling work motivated by observations of SINGS galaxies and
high redshift quasars has similarly predicted that dust growth and non-stellar
dust production can have a pronounced impact on a galaxy's dust content, while
dust destruction is less influential \citep{Draine2009, Michalowski2010,
Valiante2011, Mattsson2012, Zafar2013, Rowlands2014}.

Models without growth and without destruction produced the lowest and highest
dust-to-metal ratios for $z < 2$, respectively.  While the dust-to-metal ratio
predicted by only allowing stellar production of dust is roughly $0.3$ and
physically acceptable, it is nearly constant and displays little evolution.  In
contrast, both growth and destruction mechanisms are needed to capture the
increase in dust-to-metal ratio expected at low redshift \citep{Inoue2003a}, as
in our fiducial model and the run with low dust condensation efficiencies.  The
extent to which the dust-to-metal ratio varies within a galaxy and among
different galaxies is unclear \citep{Mattsson2012}, so it is difficult to
assess whether our fiducial dust model yields a typical dust-to-metal ratio.
In any case, from Figure~\ref{FIG:dust_ratio_comparison_Ccomp} we can conclude
that varying dust condensation efficiencies makes little difference to the
dust-to-metal ratio evolution, and while adopting a constant destruction
timescale instead of using local SNe-driven destruction leads to a
higher dust-to-metal ratio at high redshift, the effect is nowhere near as
severe as limiting dust growth or destruction.

In Figure~\ref{FIG:dust_to_gas_metal_Ccomp} we plot the dust-to-gas ratio
versus dust-to-metal ratio at $z = 2$, $1$, and $0$ for every gas cell in the
halo using the fiducial run, a model without dust growth, and a model without
dust destruction.  Given the importance of metallicity in regulating the
transition between stellar production-dominated dust evolution and ISM-led dust
growth, we also compute the mean metallicity for each region of dust-to-gas and
dust-to-metal ratios.  While the Aquarius C halo is fairly quiescent over this
redshift range, we see qualitative differences between these three dust models
with slight evolution over time.

\begin{figure*}
\centering
\includegraphics{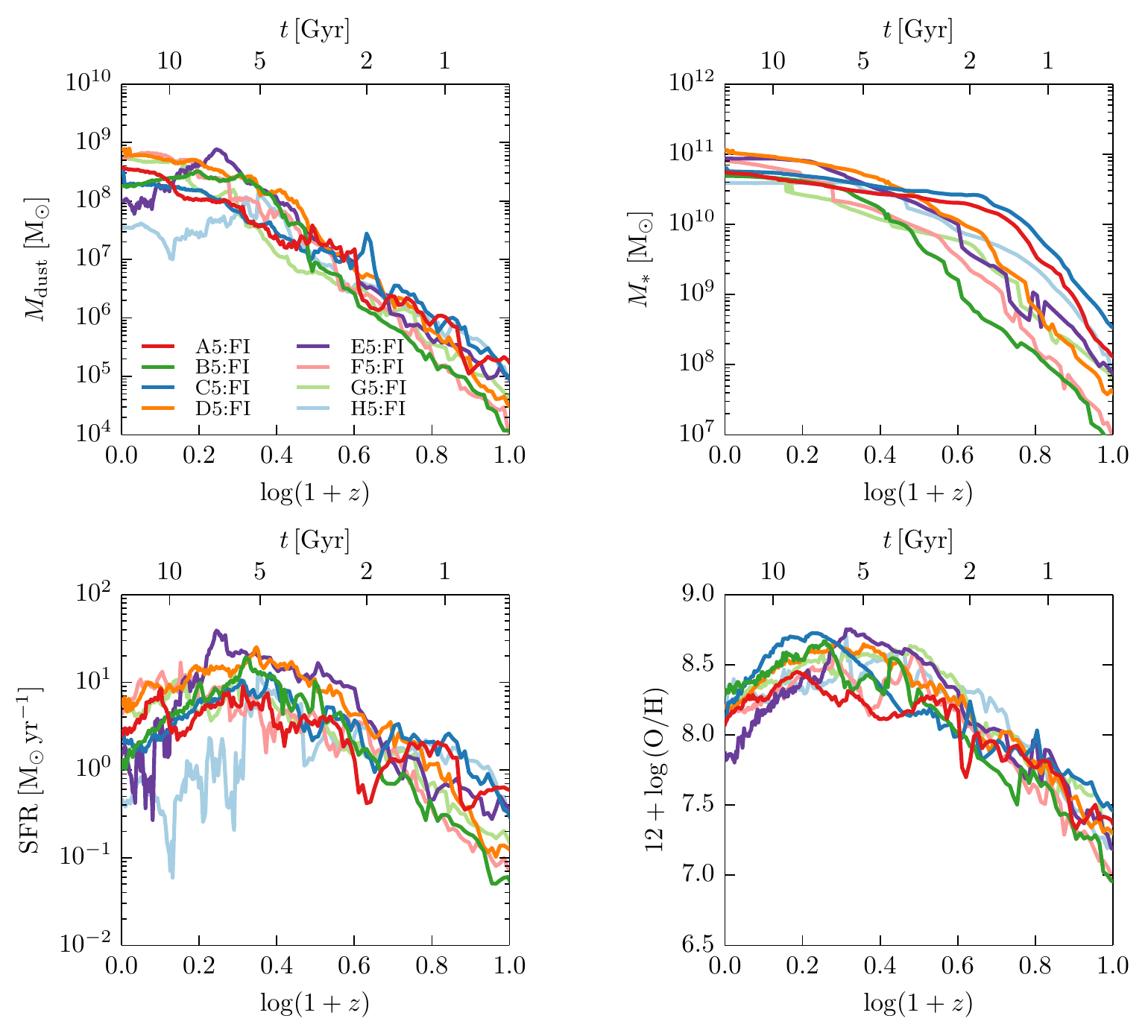}
\caption{Clockwise from top left, a comparison of dust mass, stellar mass,
gas-phase metallicity, and star formation rate for the Aquarius sample of galaxies
with standard dust model parameters.  These quantities include contributions
only from the dense galactic disc and not the hot halo.}
\label{FIG:dust_quantities_comparison_halocomp}
\end{figure*}

At all redshifts, the fiducial model contains cells across a wide range of
dust-to-metal ratio.  Regions of larger dust-to-gas ratio tended to be more
metal-rich, with the peak dust-to-gas ratio and metallicity increasing by
roughly $1 \, \text{dex}$ and $0.3 \, \text{dex}$, respectively, from $z = 2$
to $z = 0$.  In the fiducial model, low dust-to-gas ratios at $z = 0$
were concentrated in cells with low dust-to-metal ratio.  While one-zone dust
models are incapable of producing such phase space plots, previous
smoothed-particle hydrodynamical simulations have explored the relation between
dust-to-gas and dust-to-metal ratios for fiducial dust parameters
\citep{Bekki2015a}.  While our results reproduce the overall increase in
dust-to-gas ratio from $z = 2$ to $z = 0$ and roughly $1 \, \text{dex}$ scatter
in typical dust-to-gas ratio at $z = 0$, previous work finds nearly all gas
cells concentrated in a narrow band of dust-to-metal ratio.  We find
moderate variation in the dust-to-metal ratio within the galaxy, results
consistent with the visualisations in Figures~\ref{FIG:projected_densities} and
\ref{FIG:proj_Zdust_disk}.

In comparison, the models lacking growth and destruction mechanisms see less
diversity among gas cells.  The no growth model has $Z_\text{dust}/Z < 0.5$ for
essentially every cell, with the typical dust-to-metal ratio decreasing towards
$z = 0$.  Conversely, the no destruction model sees $Z_\text{dust}/Z > 0.4$ for
nearly all gas cells at $z = 1$ and $z = 0$.  Additionally, by $z
= 0$, the model without destruction witnesses a peak dust-to-gas ratio roughly
$2 \, \text{dex}$ above that obtained in the no growth model.  These results
suggest that the presence of both growth and destruction mechanisms leads to
more variation in the simulated dust content of galaxies.

\subsection{Variations in the Initial Conditions}

In the remainder of this work, we use the fiducial feedback and dust models to
investigate the evolution of dust within our sample of eight Aquarius galaxies.
Figure~\ref{FIG:dust_quantities_comparison_halocomp} shows the dust mass,
stellar mass, SFR, and gas-phase metallicity for each of these galaxies as a
function of redshift.  These quantities exclude contributions from cells
in the hot halo.  The variation in galactic dust mass is
largest at low redshift, with galaxies D and H producing dust
masses of $8 \times 10^{8} \, \text{M}_\odot$ and $3 \times
10^{7} \, \text{M}_\odot$, respectively, at $z = 0$.  Changes in the
dust content largely trace changes in the overall metal mass.  We note that two
of the galaxies with the largest dust and metal masses at high
redshift, A and C, are known to have progenitors that assemble their mass more
rapidly than the others \citep{Wang2011}.  The SFRs for these galaxies cover the
range $0.4 - 7 \, \text{M}_\odot \, \text{yr}^{-1}$ at $z = 0$, and the
galaxy that experiences the largest drop in SFR from $z=1$ to $z=0$, galaxy E,
also sees the largest decline in gas-phase metallicity.
While decreased star formation produces fewer dust grains in the ISM,
it also weakens SN-based grain destruction, ultimately enabling gas-phase
depletion.

\begin{figure*}
\centering
\includegraphics{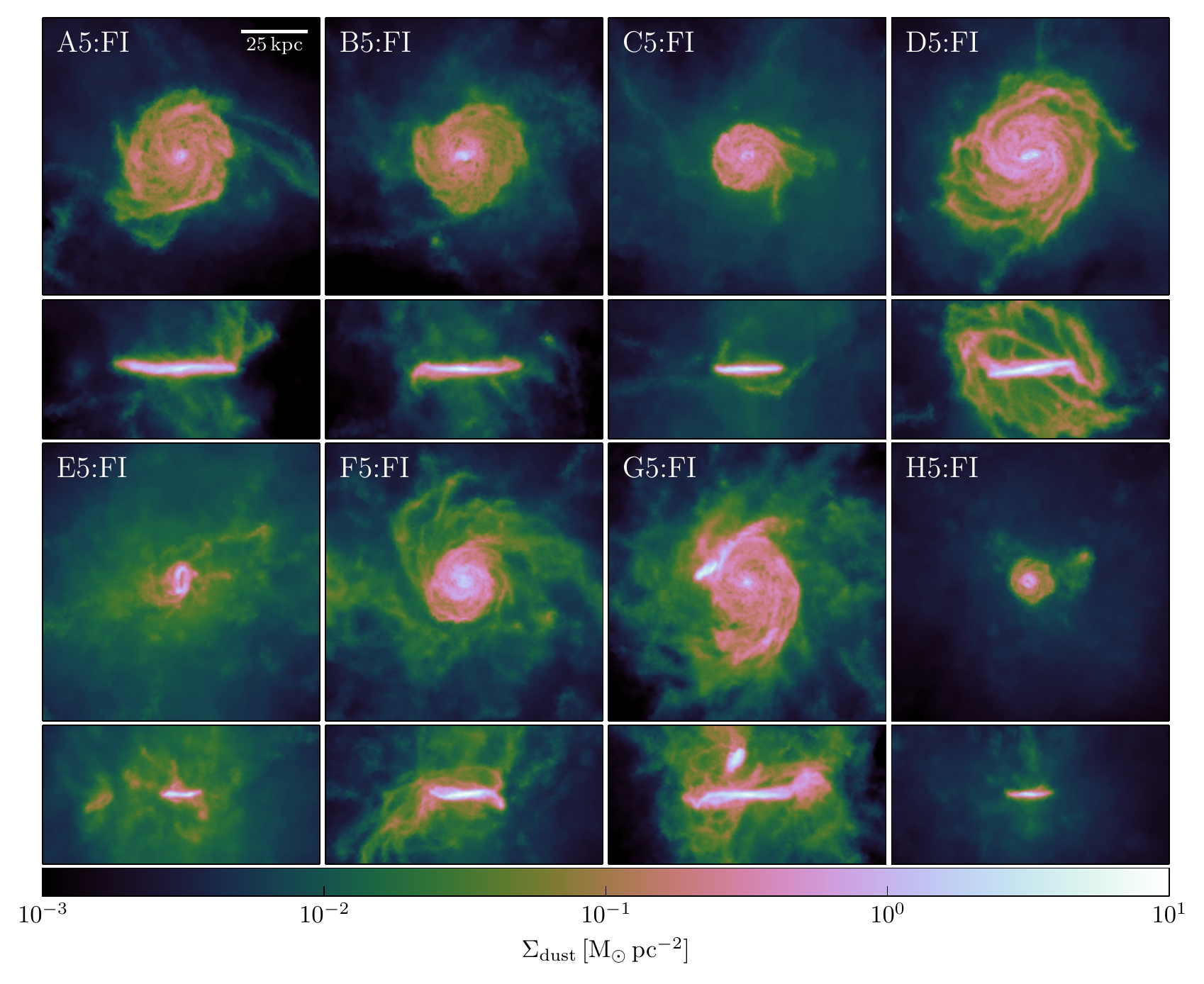}
\caption{Dust surface density maps for all eight simulated Aquarius haloes at $z
= 0$ using the fiducial model.  For each halo, face-on (top) and edge-on
(bottom) projections are shown.  The scale bar in the upper left indicates $25
\, \text{kpc}$.  Projections were performed in a cube of side length $150 \,
h^{-1} \, \text{kpc}$ centered on the halo's potential minimum.  All
haloes form galactic discs in which dust is concentrated, although the discs
vary in size.}
\label{FIG:proj_halocomp}
\end{figure*}

To further contrast these haloes, in Figure~\ref{FIG:proj_halocomp} we display
face-on and edge-on dust surface density projections for the Aquarius sample at
$z = 0$.  These haloes are at various stages of galactic disc formation.  In
some cases, as for halo C, a quiescent disc of dust has formed with an abrupt
drop in dust surface density when moving to the CGM.  Other haloes, including D,
F, and G, display much more spatial variation, especially off of the disc
plane.  In particular, halo G is perturbed by a satellite galaxy at $z = 0$
\citep{Marinacci2014}.  A dusty disc of comparable size was found in recent
smoothed-particle hydrodynamical simulations of a $10^{12} \, \text{M}_\odot$
halo \citep{Bekki2015a}, resembling our results for halo C.  Furthermore, the
diversity of dust surface densities that we see could impact the creation of
synthetic galaxy images and spectra using spectral energy distribution
modelling \citep{Silva1998, Jonsson2006, Bernyk2014, Snyder2015, Torrey2015}.

\begin{figure*}
\centering
\includegraphics{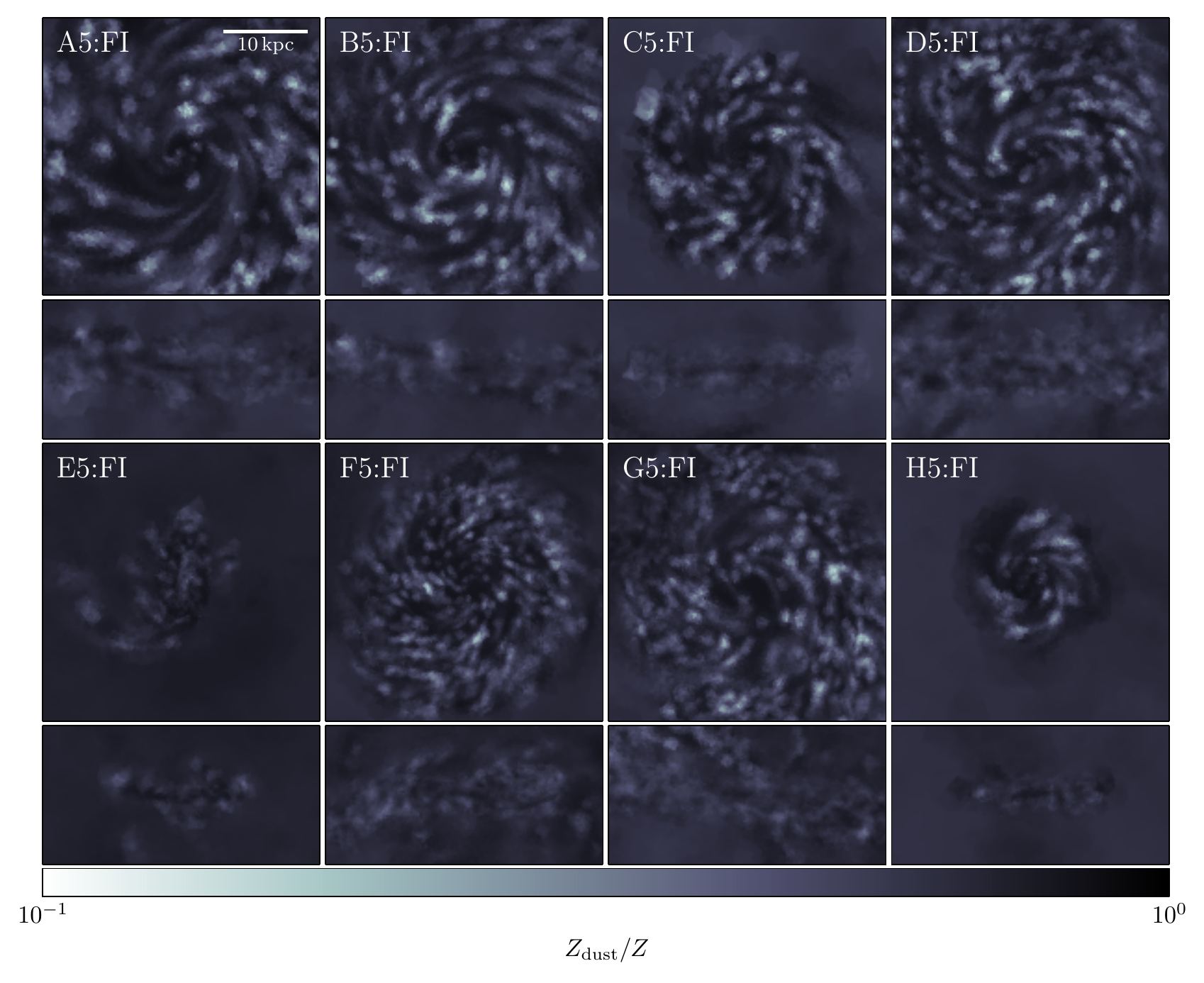}
\caption{Projected dust-to-metal ratio in the inner disc region for the
haloes shown in Figure~\ref{FIG:proj_halocomp}.  Face-on (top) and edge-on
(bottom) projections are displayed for each halo column.  The projection volume
was a cube of side length $50 \, h^{-1} \, \text{kpc}$.
Figure~\ref{FIG:proj_Zdust_disk} displayed higher resolution images for the
Aquarius C halo at $z = 2$, $1$, and $0$.}
\label{FIG:proj_halocomp_Zdust}
\end{figure*}

Figure~\ref{FIG:proj_halocomp_Zdust} shows the projected dust-to-metal ratio of
these haloes at $z = 0$ and indicates that the dust-to-metal ratio exhibits
fluctuations within the galactic disc.  Analytical arguments suggest
that dust-to-metal gradients can be used to estimate the relative strength of
dust growth and destruction in the ISM, with flat dust-to-metal profiles
expected in the absence of dust growth and destruction \citep{Mattsson2012}.
Rather compact galaxies, as in haloes E and H, see less spatial
variation in the dust-to-metal ratio than the larger galactic discs in
Figure~\ref{FIG:proj_halocomp}, which tend to have more pockets of low
and high dust-to-metal ratio.  Stellar density projections for the Aquarius
suite have previously been made using the same physical model without dust
\citep{Marinacci2014} and show that depressions in dust-to-metal ratio
are similar in size to regions of high stellar density.  In
particular, haloes A, C, F, and G have some of the largest stellar
discs and also some of the most variation in dust-to-metal ratio.
While our dust model does not properly account for thermal sputtering and
grain-grain collisions in hot regions of the CGM,
Figure~\ref{FIG:proj_halocomp_Zdust} suggests the presence of noticeable
dust-to-metal ratio fluctuations in Milky Way-mass galactic discs that
trace stellar structure.

\begin{figure*}
\centering
\includegraphics{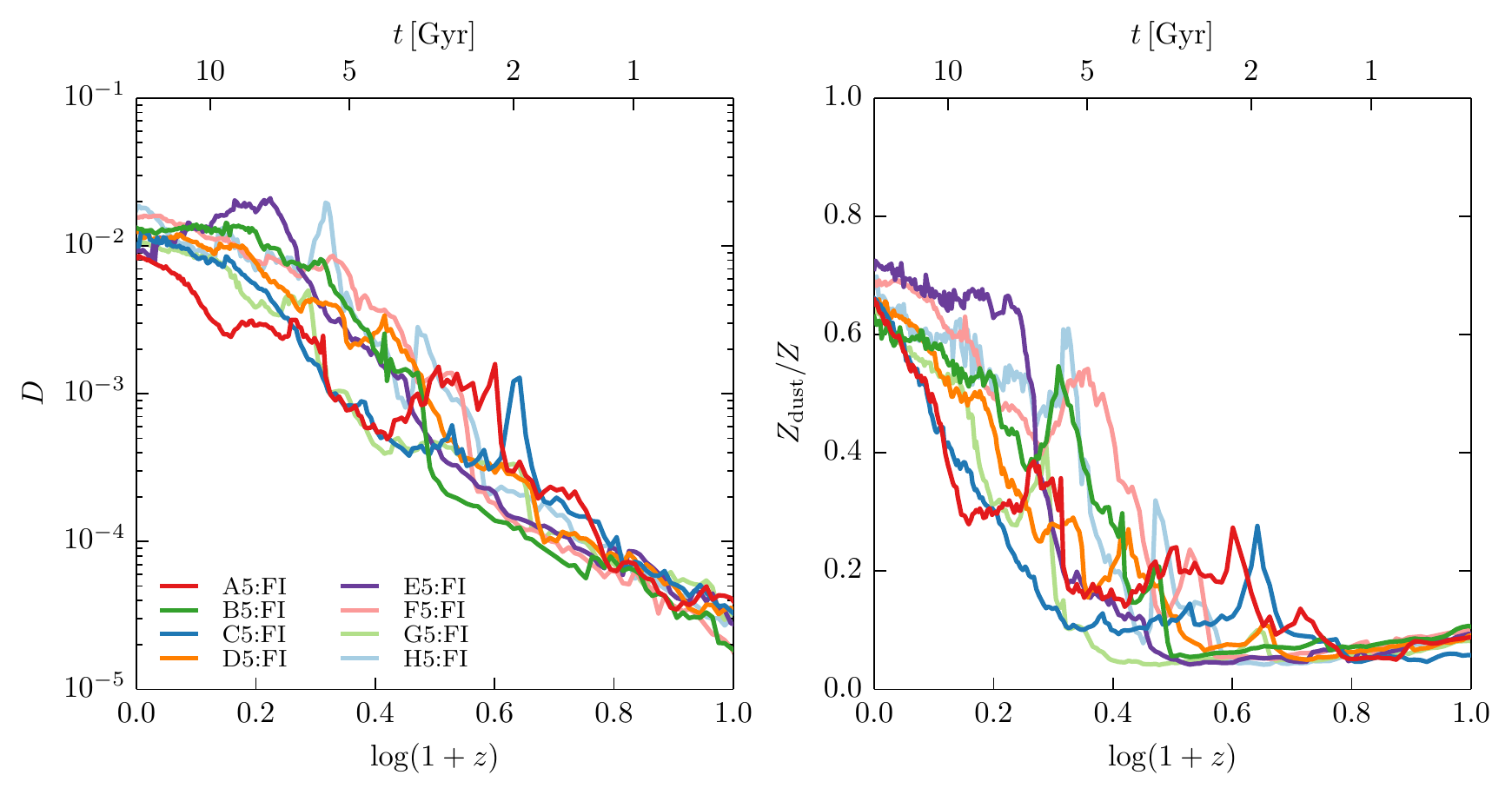}
\caption{Comparison of dust-to-gas ratio ($D$) and dust-to-metal ratio
($Z_\text{dust}/Z$) for a variety of Aquarius galaxies with standard dust model
parameters.  The increase in dust-to-gas ratio can be as great as an order of
magnitude from $z=2$ to $z=0$.  There is significant variation in dust-to-metal
ratio among the haloes at high redshift, though all settle near
$Z_\text{dust}/Z \approx 0.65$ at $z = 0$.}
\label{FIG:dust_ratio_comparison_halocomp}
\end{figure*}

\begin{figure*}
\centering
\includegraphics{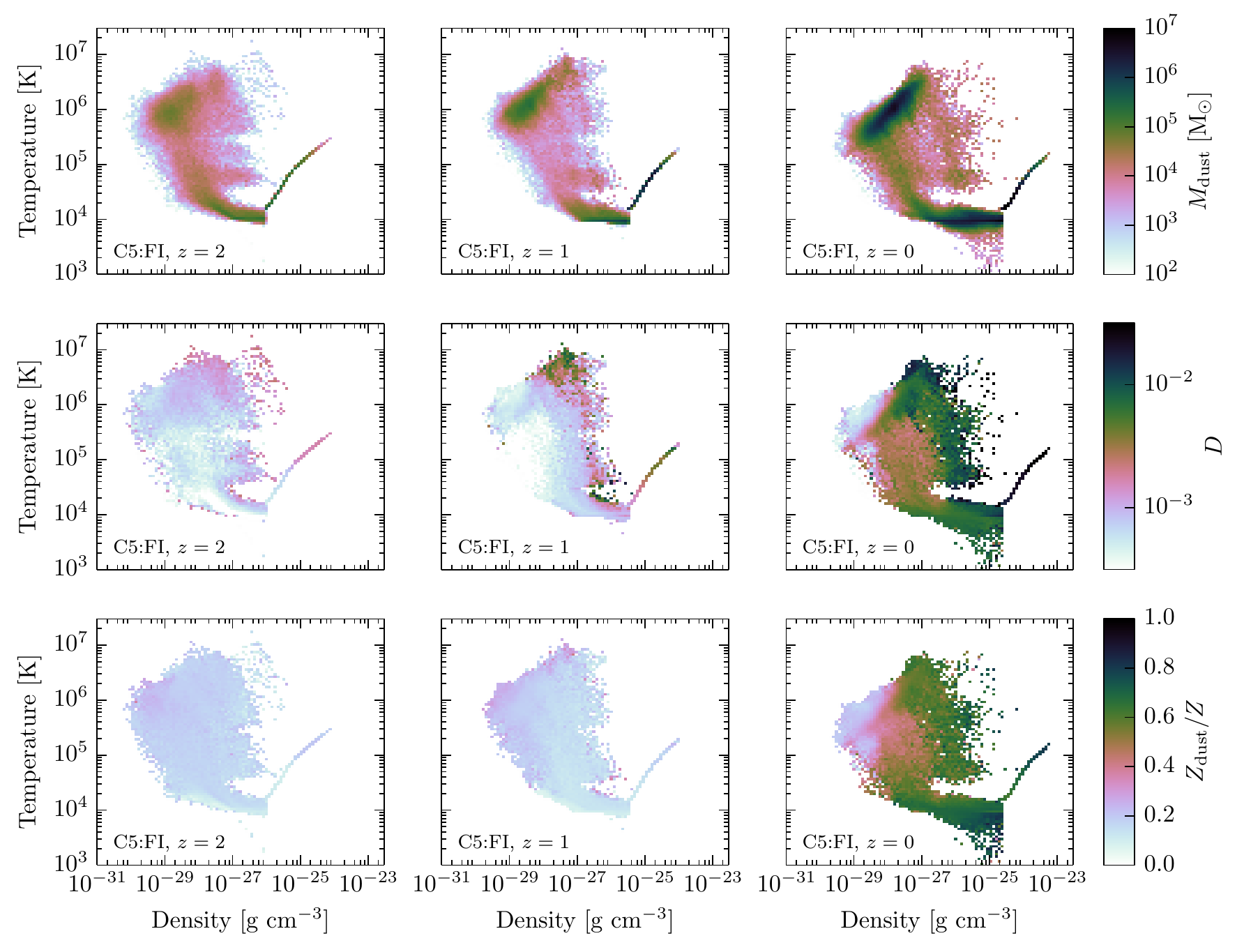}
\caption{Phase diagram of gas temperature versus density for all gas cells in
the Aquarius C halo at $z = 2$, $1$, and $0$ (left, middle, and right columns,
respectively) where bins are colored according to the total dust mass of
constituent gas cells (top row), mass-weighted dust-to-gas ratio (middle), and
mass-weighted dust-to-metal ratio (bottom).  Each row adopts its own color
scale, and fiducial dust model parameters were used.  The high-density region
where gas cools is governed by an equation of state.  At $z = 0$, the
dust-to-gas and dust-to-metal ratios are smallest in low density gas.}
\label{FIG:dust_Trho_phase}
\end{figure*}

To investigate the dust-to-metal ratio evolution more quantitatively, in
Figure~\ref{FIG:dust_ratio_comparison_halocomp} we plot the dust-to-gas and
dust-to-metal ratios as a function of redshift for these galaxies.
The dust-to-gas ratios tend to evolve in similar fashion, reaching Milky
Way-like values of $(0.8 - 1.8) \times 10^{-2}$ by $z = 0$.  At $z =
7.5$, where A1689-zD1 has been observed to have a near-Galactic dust-to-gas
ratio \citep{Watson2015}, the Aquarius suite of galaxies sees
dust-to-gas ratios slightly under $10^{-4}$.  Our results suggest that
A1689-zD1 and other dusty high-redshift galaxies do not arise from Milky
Way-like progenitors and motivate further study of the dust-to-gas ratio over a
range of halo masses.  Recent models predict that larger haloes see more
dramatic growth in the dust-to-gas ratio towards low redshift and that smaller
systems have less dynamic dust-to-gas ratios \citep{Bekki2015a}.  A larger
statistical sample of galaxies would also enable comparison to observed dust
mass functions \citep{Dunne2000, Dunne2011}.

The dust-to-metal ratios in Figure~\ref{FIG:dust_ratio_comparison_halocomp}
display more diversity, with $z = 0$ values ranging from roughly $0.6$
to $0.7$.  While high depletion has been observed for some elements,
including Si, Mg, and Fe, in the local interstellar cloud \citep{Kimura2003},
this wouldn't greatly change the overall dust-to-metal ratio.
Estimates of the dust-to-metal ratio for typical galaxies, including
the Milky Way and Magellanic Clouds, lie closer to $0.5$ \citep{Aguirre1999,
DeCia2013}.  All show an increase in dust-to-metal ratio from $z = 2$ to $z = 0$,
expected of galaxies \citep{Inoue2003a}.  Before $z \approx 3$, about
half of galaxies display a flat dust-to-metal ratio near $0.1$.
However, there is some variation in when the Aquarius galaxies see the most
growth in dust-to-metal ratio.  Galaxies A and C, which assemble the
majority of their mass more quickly than the others \citep{Wang2011}, show some
of the earliest growth in dust-to-metal ratio, suggesting that accretion
history may influence dust-to-metal ratio evolution.

In Figure~\ref{FIG:dust_Trho_phase}, we construct temperature-density phase
diagrams for all gas cells within halo C at $z = 2$, $1$, and $0$.  We analyze
gas cells in three different ways: by total dust mass, by mass-weighted
dust-to-gas ratio, and by mass-weighted dust-to-metal ratio.  We note that from
$z = 2$ to $z = 0$ dust mass becomes increasingly concentrated in hot, diffuse
halo gas with $T \sim 10^{6} \, \text{K}$ and dense, star-forming ISM gas with
$T \sim 10^{4} \, \text{K}$ near the equation of state transition.  By
contrast, the phase diagram region with $T \sim 10^{5} \, \text{K}$ has a total
dust mass largely unchanged with redshift.  While SDSS data do suggest that the
dust masses of galactic haloes are similar to those found in galactic discs
\citep{Menard2010}, the results in Figure~\ref{FIG:dust_Trho_phase} also
motivate better modelling of dust evolution in the diffuse halo where grains
may travel at higher velocities, have lower sticking efficiencies when
impacting gas-phase metals, and undergo grain-grain collisions.  Thus, it is
likely that the halo dust mass is overstated, especially at $z = 0$, but we
still expect the presence of dust beyond the galactic disc.

Dust-to-gas ratios tend to increase from $z = 2$ to $z = 0$
most rapidly in high-density regions.  Gas density correlates
more strongly with average dust-to-gas ratio than does temperature.  We see a
similar trend in dust-to-metal ratio, as dense gas above the equation of state
transition sees its dust-to-metal ratio increase from roughly $0.2$ to $0.6$
over this redshift range.  In contrast, low-density gas with $\rho \sim
10^{-29} \, \text{g} \, \text{cm}^{-3}$ barely changes in dust-to-metal ratio.
This result is physically justified, since the dust growth timescale is
shortest in high-density gas, and agrees with observations of elemental
depletions increasing with gas density \citep{Jenkins2009}.

\begin{figure}
\centering
\includegraphics{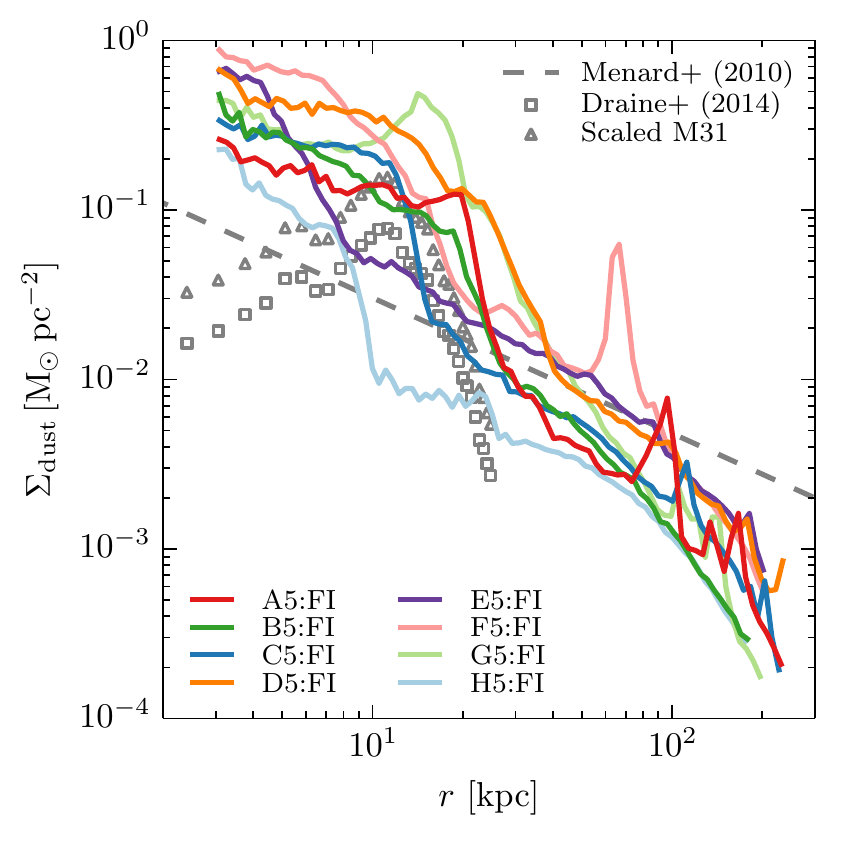}
\caption{Radial profiles of dust surface density ($\Sigma_\text{dust}$) in the
disc plane for all simulated haloes, plotted at $z = 0$.  The radial profiles
extend out to the virial radius of each halo as defined in Table~\ref{TAB:ICs}.
We compare with the observed dust profile for M31 \citep[gray
squares;][]{Draine2014} and an M31 profile scaled by a factor of two (gray
triangles), given that the Galactic dust content may vary from that of M31.
The gray dashed line shows the $\Sigma_\text{dust} \propto r^{-0.8}$ scaling
observed in SDSS data \citep{Menard2010}, with normalisation adjusted to the
Aquarius data.}
\label{FIG:dust_radial_density_halocomp}
\end{figure}

In Figure~\ref{FIG:dust_radial_density_halocomp}, we display the radial dust
surface density profiles for each of the Aquarius haloes at $z = 0$.
There is fairly little evolution from $z = 1$ to $z = 0$, although the growth
in virial radius causes $\Sigma_\text{dust} \sim 10^{-3} \, \text{M}_\odot \,
\text{pc}^{-2}$ out to $r \approx 200 \, \text{kpc}$ by $z = 0$.  These Milky
Way-sized haloes have dust profiles fairly consistent with observations of M31,
with higher normalisations in the inner disc.  The Aquarius haloes
tend to show rather sharp drops in dust surface density when moving outside of
the galactic disc: for example, halo C witnesses a decline of almost an order
of magnitude in $\Sigma_\text{dust}$ near $r \approx 20 \, \text{kpc}$,
although it is not as sharp as the drop seen for M31.  The Aquarius haloes have
dust profiles that decay from $r \approx 20 \, \text{kpc}$ to $r \approx 100 \,
\text{kpc}$ in rough agreement with the observed scaling $\Sigma_\text{dust}
\propto r^{-0.8}$ seen in SDSS data \citep{Menard2010}.

\begin{figure}
\centering
\includegraphics{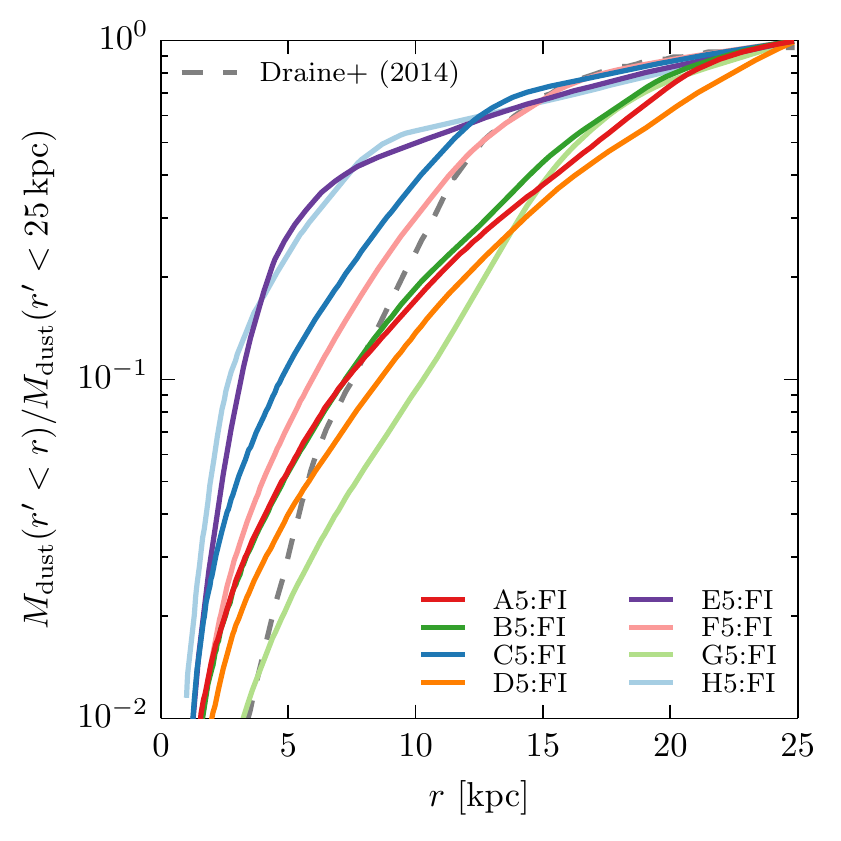}
\caption{Enclosed dust mass as a function of radial distance for all haloes at
$z = 0$, normalised to $M(r' < 25 \, \text{kpc})$, the mass within $25 \,
\text{kpc}$ of the disc center.  The normalised cumulative dust mass
distribution observed for M31 is given by the gray dashed line
\citep{Draine2014}.  While there is some scatter in the Aquarius haloes, the
enclosed dust mass profiles are similar to that of M31.}
\label{FIG:dust_radial_Menc_halocomp}
\end{figure}

In Figure~\ref{FIG:dust_radial_Menc_halocomp} we compare the cumulative dust
mass distributions for the Aquarius galaxies with the observed distribution for
M31 \citep{Draine2014}, normalising to the total dust mass within a radial
distance of $25 \, \text{kpc}$ from galactic center.  We do not necessarily
expect the Aquarius suite to have cumulative dust mass distributions identical
to that of M31 but use these observational data for guiding purposes.
The Aquarius haloes witness a slightly steeper enclosed dust mass distribution
for $r < 10 \, \text{kpc}$ than is observed for M31, but the simulated and
observed distributions are in rough agreement.  In particular, we find
in the Aquarius sample that the dust masses enclosed within $10 \, \text{kpc}$
are between 9 and 54 per cent of the total dust mass within $25 \,
\text{kpc}$.  For comparison, the observed value for M31 is about 25 per cent
\citep{Draine2014} and recent smoothed-particle hydrodynamical simulations
using a Milky Way-sized halo with dust growth mechanism similar to ours found
that roughly 18 per cent of the dust mass inside of $25 \, \text{kpc}$ is
contained within the innermost $10 \, \text{kpc}$ \citep{Bekki2015a}.

\subsection{Scaling Relations}

To better understand how realistically our sample of eight Aquarius galaxies
produce dust, we compare against a number of empirical scaling relations seen
in recent observations \citep{Draine2007, Galametz2011, Kennicutt2011,
Corbelli2012, Cortese2012, RemyRuyer2014}.  These scalings enable valuable
predictions about the dust content in very metal-poor galaxies
\citep{Fisher2014} and high-redshift, dusty quasars \citep{Fan2003,
Bertoldi2003, Valiante2009}.  Also, these relations help determine the
quantities that most effectively trace the presence of dust in galaxies.  One
of the most prominent of these scalings is the dust-metallicity relation, which
finds a dust-to-gas ratio that scales fairly linearly with metallicity
\citep{Draine2007, Galametz2011, Kennicutt2011, RemyRuyer2014} and is often
used to estimate the dust mass within a galaxy.  It is therefore interesting to
consider such scaling relations and to investigate any variations in our
Aquarius sample.  In the observational comparisons below, we compute all
quantities using dense ISM gas, as determined by the cut in temperature-density
phase space of all gas cells belonging to the main halo group given by
Equation~(\ref{EQN:Torrey2012}).  This cut isolates the galactic disc and thus
ignores the effect of dust growth in the CGM that is possibly too strong.

\begin{figure}
\centering
\includegraphics{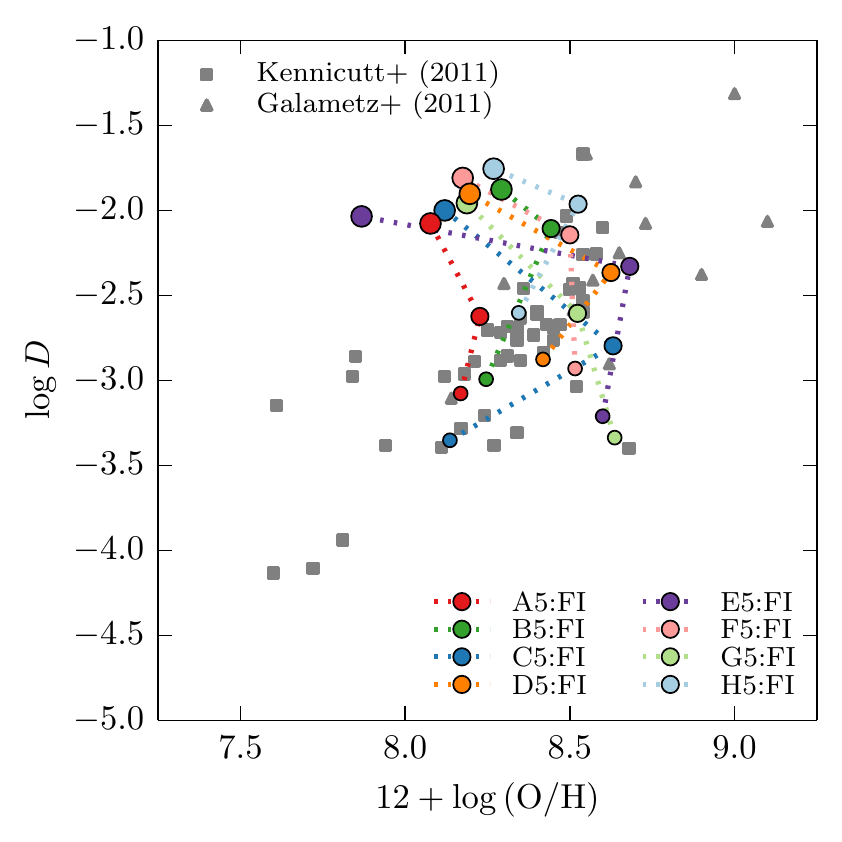}
\caption{Dust-to-gas ratio ($D$) versus gas-phase metallicity for the Aquarius
haloes plotted at $z = 2$, $1$, and $0$, with smaller circles denoting higher
redshift.  The same observational data of local galaxies from
\citet{Kennicutt2011} and \citet{Galametz2011} as in
Figure~\ref{FIG:dust_metallicity_feedback} are shown in gray.  These galaxies
all show decreases in gas-phase metallicity from $z = 1$ to $z = 0$ but fall
close to the observed scatter at $z = 2$ and $z = 1$.}
\label{FIG:dust_metallicity_halocomp}
\end{figure}

We first show the dust-metallicity relation for the Aquarius galaxies in
Figure~\ref{FIG:dust_metallicity_halocomp}, plotting dust-to-gas ratio versus
gas-phase metallicity at $z = 2$, $1$, and $0$, with the gas-phase metallicity
calculated from oxygen and hydrogen abundances.  As in
Figure~\ref{FIG:dust_metallicity_feedback}, we compare against observational
data from \citet{Kennicutt2011} and \citet{Galametz2011}.  Previous models have
been fairly successful at reproducing the dust-metallicity relation
\citep{Dwek1998, Lisenfeld1998, Calura2008, Bekki2015a, Feldmann2015} and have
investigated the nature and scatter of this relation at low metallicity
\citep{RemyRuyer2014}.  In comparison to several models that see a strictly
monotonic increase in dust-to-gas ratio and gas-phase metallicity towards low
redshift \citep{Lisenfeld1998, Bekki2015a, Feldmann2015}, the Aquarius haloes
display more diverse redshift evolution while remaining in agreement with the
observed dust-metallicity relation at $z = 2$ and $z = 1$.  While the
data for $z = 0$ capture the positive correlation between dust-to-gas ratio and
gas-phase metallicity, they tend to lie above the observed relation, as in
Figure~\ref{FIG:dust_metallicity_feedback}.

While some haloes, like A and B, display small drops in
gas-phase metallicity from $z = 1$ to $z = 0$, others show much more pronounced
behaviour. In particular, haloes C and E have some of the largest
declines in SFR from $z = 1$ to $z = 0$ in our sample, weakening grain
destruction and enhancing depletion of gas-phase metals.  Given that the
temperature-density cut used to isolate dense gas in the galactic disc will
minimise the influence of dust in the CGM,
Figure~\ref{FIG:dust_metallicity_halocomp} suggests that the dust-metallicity
relation can evolve in a non-monotonic fashion even near the galactic center.
However, the behaviour at $z = 0$ together with fairly high
dust-to-metal ratios in Figure~\ref{FIG:dust_ratio_comparison_halocomp}
indicate that gas-phase metals may be too heavily depleted at low redshift.
Additionally, while there is evidence for a dust-metallicity relation that is
less steep for metallicities below 8.0 \citep{RemyRuyer2014}, suggesting that
the global dust-metallicity trend cannot be fit by a single power law, the
Aquarius galactic disc gas-phase metallicities lie above this region.  Future
simulations of full cosmological volumes will allow us to explore the
dust-metallicity scaling at low metallicity.

\begin{figure}
\centering
\includegraphics{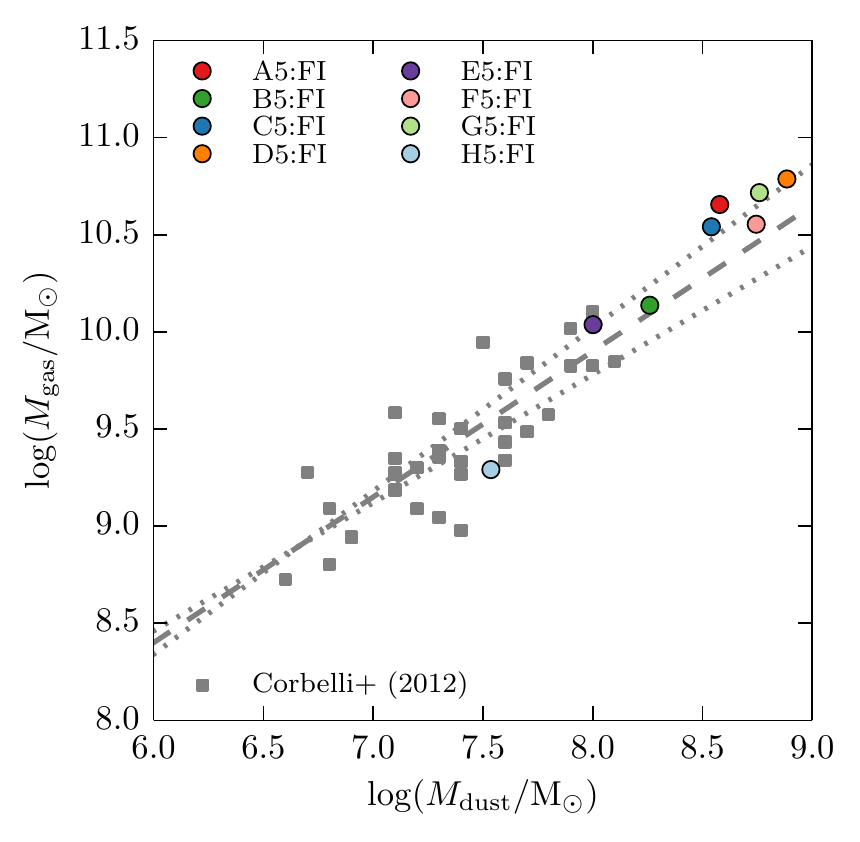}
\caption{Scaling relation between dust mass ($M_\text{dust}$) and gas mass
($M_\text{gas}$) for the eight Aquarius haloes, with observational data from the
Herschel Virgo Cluster Survey \citep{Corbelli2012}.  The dashed line represents
the best fit scaling for the sample of 35 metal-rich galaxies in Virgo at $z =
0.003$, with $M_\text{gas} \propto M_\text{dust}^{0.75}$.  Uncertainties in the
regression parameters yield scaling fits given by the dotted lines.  The
Aquarius haloes are more massive than those in this observational sample and
display a slightly steeper gas-dust scaling.}
\label{FIG:dust_gas_scaling_halocomp}
\end{figure}

In Figure~\ref{FIG:dust_gas_scaling_halocomp}, we compare the dust mass-gas
mass scaling seen for our Aquarius suite with observations of metal-rich
galaxies from the Herschel Virgo Cluster Survey \citep{Corbelli2012}.  These
observational data found the best fit scaling $M_\text{gas} \propto
M_\text{dust}^{0.75}$, suggesting that the dust-to-gas ratio increases weakly
with gas mass.  That is, the presence of additional gas leads to increased star
formation, which in turn produces enough dust to cause the dust-to-gas ratio to
rise.  We find that the Aquarius data obey a slightly steeper scaling of gas
mass with dust mass, consistent with the results in
Figure~\ref{FIG:dust_ratio_comparison_halocomp} showing fairly little variation
in the dust-to-gas ratio in our sample.  However, the Aquarius haloes tend to
be more gas- and dust-rich than those from the Herschel Virgo Cluster Survey,
and future work should investigate whether our observed scaling holds for lower
mass systems.

\begin{figure}
\centering
\includegraphics{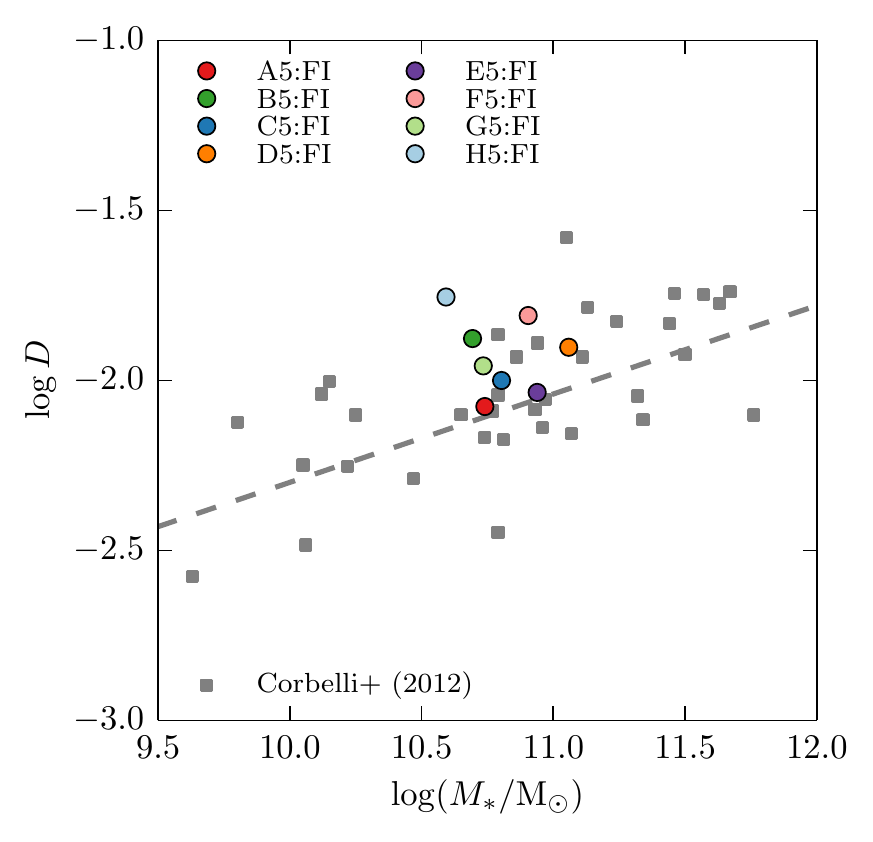}
\caption{Dust-to-gas ratio ($D$) plotted as a function of stellar mass ($M_*$)
for our sample of Aquarius haloes, with the same Herschel Virgo Cluster Survey
data as in Figure~\ref{FIG:dust_gas_scaling_halocomp} given in gray
\citep{Corbelli2012}.  The observed weak scaling with $D \propto M_*^{0.26}$
is indicated by the dashed line.  The scatter of Aquarius haloes is similar
to that in these observations.}
\label{FIG:dust_D_scaling_halocomp}
\end{figure}

The Herschel Virgo Cluster Survey also investigated the relationship between
dust-to-gas ratio and stellar mass, observing the weak scaling $D \propto
M_*^{0.26}$.  This scaling is possibly weak because increased star formation
injects more dust into the ISM through stellar production but also leads to SNe
destroying dust more rapidly.  In Figure~\ref{FIG:dust_D_scaling_halocomp}, we
compare the $D-{M_*}$ data from the Aquarius suite with the Herschel scaling
and find the Aquarius data to lie close to the observed best fit.  While the
Aquarius haloes show a scatter in dust-to-gas ratio that roughly matches that
seen in the Herschel data, our haloes do not cover the full range of stellar
masses observed by Herschel.

\begin{figure}
\centering
\includegraphics{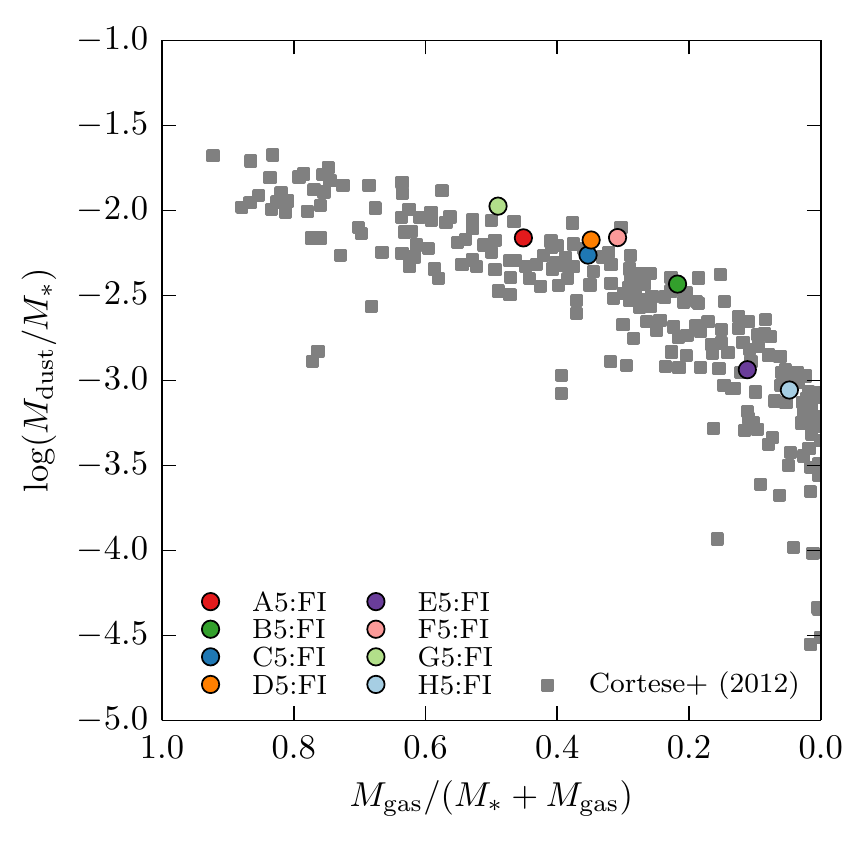}
\caption{Relation between dust-to-stellar mass ratio ($M_\text{dust}/M_*$) and
gas fraction ($M_\text{gas} / (M_* + M_\text{gas})$) for all simulated Aquarius
haloes, shown as colored circles, with observational data from the Herschel
Reference Survey at $z \leq 0.006$ provided as gray squares
\citep{Cortese2012}.  The observational data include both galaxies classified
as H$\,$\textsc{i}-normal and H$\,$\textsc{i}-deficient and omit upper limits.
Both simulation and observation predict that galaxies with larger gas fraction
have higher dust-to-stellar mass ratio.}
\label{FIG:dust_MdustMstar_gasfrac_halocomp}
\end{figure}

In Figure~\ref{FIG:dust_MdustMstar_gasfrac_halocomp}, we compare the observed
scaling between dust-to-stellar mass ratio and gas fraction ($M_\text{gas} /
(M_* + M_\text{gas})$) seen in Herschel Reference Survey data
\citep{Cortese2012} with results from our Aquarius sample.  The dust-to-stellar
mass ratio is expected to anti-correlate with stellar mass and decrease from
late- to early-type galaxies \citep{Cortese2012}.  Late-type galaxies with high
gas fraction see more efficient stellar injection of dust, while early-type
galaxies are more influenced by grain destruction.  The Aquarius suite is in
good agreement with the observed dust-to-stellar mass ratio scaling, covering a
range of gas fractions from less than $0.1$ to $0.5$.

\begin{figure}
\centering
\includegraphics{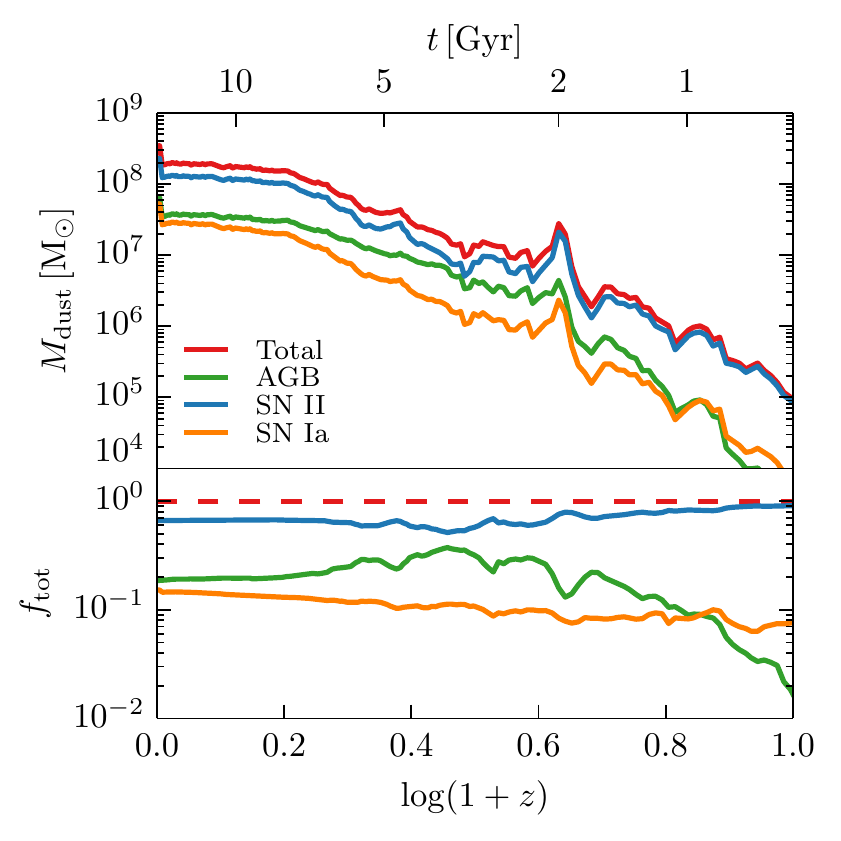}
\caption{Evolution of dust mass contributions from different stellar types for
the Aquarius C galaxy using the fiducial dust model.  The bottom subpanel shows
the fraction of total dust mass ($f_\text{tot}$) contributed by each component.
As noted in Section~\ref{SEC:dust_destruction}, the relative proportions of
dust produced by AGB stars and SNe are kept constant during changes in dust
mass from dust growth or destruction.  Type II SNe are responsible for roughly
80 per cent of dust formed in the first Gyr and two-thirds of dust at $z = 0$.
This figure captures the delayed production of dust by AGB stars owing to their
long lifetimes.}
\label{FIG:dust_channels}
\end{figure}

While the Aquarius haloes compare favorably to these dust scaling relations at
$z = 0$, in the future we would like to perform similar analyses for multiple
epochs.  There are estimates of the dust mass function for $z < 0.5$
\citep{Dunne2011} and for high-redshift submillimetre sources
\citep{Dunne2003a}.  The relation between dust mass, SFR, and stellar mass has
also been studied out to $z = 2.5$ \citep{Santini2014}.  These scalings would
be interesting in part because our Aquarius haloes displayed more diversity at
high redshift than at $z = 0$.

Especially at high redshift, it is important to understand the relative dust
production strengths of SNe and AGB stars, since this helps determine whether
dusty galaxies and quasars could have dust content driven by stellar sources as
opposed to interstellar dust growth.  Chemical evolution models suggest that
AGB stars played an important role in forming the large dust mass of SDSS
J1148+5251 at $z = 6.4$, possibly contributing over half of the observed dust
\citep{Valiante2009}.  Observations of dust in the ejecta of SNe fall short of
the $1 \, \text{M}_\odot$ or more a SN needs to produce in order to explain the
dust content of SDSS J1148+5251 \citep{Todini2001, Sugerman2006, Dwek2007,
Lau2015}, and if SNe are not dominant producers of dust, high SFRs might be
needed to form dusty $z \gtrsim 5$ galaxies \citep{Morgan2003, Santini2014}.
Given that the progenitors of AGB stars take longer to evolve off of the main
sequence than SNe, it is unclear whether AGB stars can have a large impact at
high redshift.  In Figure~\ref{FIG:dust_channels}, we compare the dust mass
contributions of SNe Ia, SNe II, and AGB stars for the Aquarius C galaxy as a
function of redshift.  We see that 80 per cent or more of dust originated in
SNe II for $z \gtrsim 5$, with the contribution from AGB stars peaking at
roughly 40 per cent for $z \approx 2$.  Recent investigation of the Small
Magellanic Cloud suggests that even at low redshift AGB stars are not dominant
producers of dust \citep{Boyer2012}.  In this Aquarius galaxy, we see that by $z
= 0$ about 20 per cent of dust has its origins in AGB stars.  While future
work is needed to investigate how these stellar contributions vary with galaxy
stellar mass and metallicity, in our sample of Milky Way-sized galaxies we find
SNe to be the dominant producers of dust.

\section{Discussion and Conclusions}\label{SEC:discussion}

We have implemented a first very basic dust model in the moving-mesh code
\textsc{arepo}, adding to the existing galaxy formation physics.  Our dust
model accounts for local stellar production of dust, growth in the ISM due to
accretion of gas-phase metals, destruction via SN shocks, and dust driven by
stellar feedback winds.  Dust is also passively advected between gas cells.  We
track dust in individual chemical species as well as follow contributions from
AGB stars, SNe Ia, and SNe II.

Using this dust model, we performed cosmological zoom-in simulations of the
Aquarius suite of eight haloes with the goal of understanding how dust forms in
Milky Way-sized systems.  After investigating the effect of feedback and the
strength of various dust model components, we used the full sample of Aquarius
haloes to compare to a range of dust observations at low redshift.  We summarise
our main findings as follows.

\begin{figure}
\centering
\includegraphics{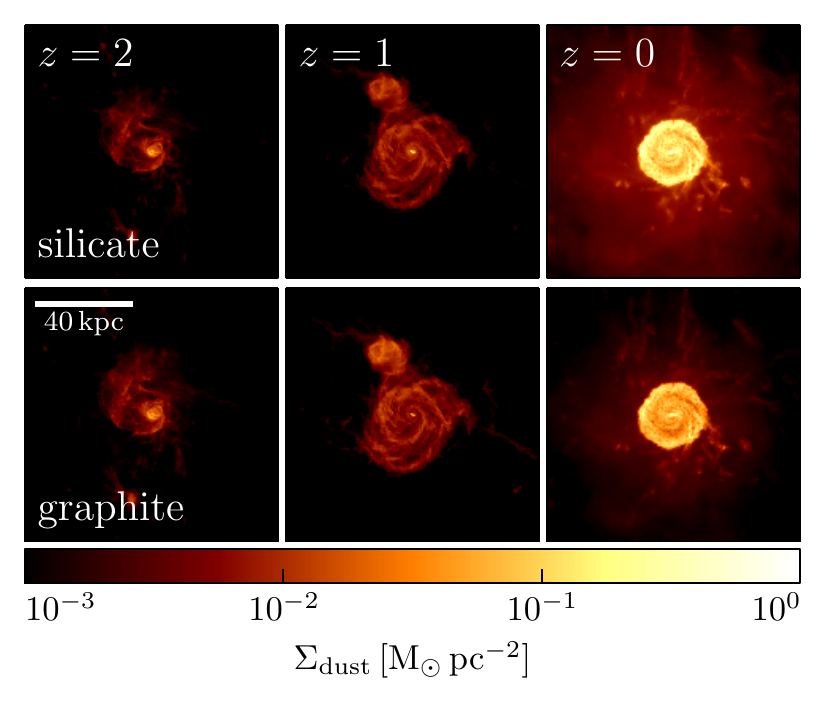}
\caption{Surface densities of silicate (top) and graphite (bottom) components
of dust for the Aquarius C halo at $z = 2$, $1$, and $0$ (left, middle, and
right columns, respectively).  The details of the projections are the same as
those in Figure~\ref{FIG:projected_densities}.  The silicate and graphite dust
components largely trace one another, although future work accounting for
radiation pressure and grain size distributions may yield more diversity.}
\label{FIG:proj_dust_type}
\end{figure}

\begin{enumerate}
\item Variations in stellar and AGN feedback impact the predictions made by our
dust model.  The absence of any feedback led to a $z = 0$ dust-to-metal ratio
of $0.2$ for one Aquarius galaxy, roughly a factor of three
lower than the value predicted by our fiducial feedback model.  Excluding
feedback led to high star formation and more efficient supernova-based dust
destruction.  On the other hand, a model with fast stellar feedback-driven
winds helped sweep dust away from the galactic center and enhance depletion of
gas-phase metals due to weakened star formation, raising the halo-wide
dust-to-metal ratio to over $0.7$.

\item Allowing gas-phase metals to deplete onto dust grains and not contribute
to metal-line cooling has the potential to affect gas-phase
metallicity.  We see a difference in halo gas-phase metallicity between runs
with and without dust of roughly $0.5 \, \text{dex}$ at $z = 0$.  This
result also motivates the future inclusion of dust cooling channels.

\item In the absence of dust growth, dust-to-gas and dust-to-metal ratios are
suppressed by more than an order of magnitude at $z = 0$ from the values
predicted by our fiducial model with dust growth and destruction mechanisms.
Without growth, we obtain a dust-to-gas ratio of $3 \times 10^{-4}$
for one Milky Way-like galaxy, below the Galactic value of roughly $10^{-2}$, and
a dust-to-metal ratio $Z_\text{dust}/Z \lesssim 0.05$.  The differences between
the fiducial model and no dust growth model are noticeable for $z
\lesssim 5$.

\item For $z \gtrsim 5$, the dust mass and dust-to-gas ratio produced by a model
adopting a constant dust destruction timescale of $\tau_\text{d} = 0.5 \,
\text{Gyr}$ are largely indistinguishable from those produced by a model
with no dust destruction.  By $z = 0$, the constant destruction
timescale model yields results similar to those from the fiducial model with
local SNe-based dust destruction.  At $z = 0$, the model with constant
destruction timescale produces a dust mass $M_\text{dust} = 3 \times 10^{8} \,
\text{M}_\odot$ in the galactic disc with a dust-to-gas ratio $D = 10^{-2}$.

\item The Aquarius haloes form dusty galactic discs with typical surface
densities $\Sigma_\text{dust} \sim 10^{-1} \, \text{M}_\odot \,
\text{pc}^{-2}$.  The inclusion of thermal sputtering and grain-grain
collisions would likely reduce the dust-to-metal ratios seen in the CGM.
Dust-to-gas ratios for these Milky Way-like galaxies increase by roughly an
order of magnitude from $z = 3$ to $z = 0$ and cover the range $(0.8 - 1.8)
\times 10^{-2}$ at $z = 0$.

\item The predicted ISM dust content of the Aquarius haloes is consistent with
a number of observed scaling relations at $z = 0$, including scalings between
dust mass and gas mass, dust-to-gas ratio and stellar mass, and dust-to-stellar
mass ratio and gas fraction.  While the overall trend of the dust-metallicity
relation is reproduced over $8.1 < 12 + \log(\text{O}/\text{H}) < 9.0$
at $z = 2$ and $z = 1$, the Aquarius galaxies do not evolve along
strictly monotonic tracks from $z = 2$ to $z = 0$.

\item Our dust model tracks contributions from individual chemical species, but
additional physics is needed to capture differences between silicate and
graphite grain types.  As shown in Figure~\ref{FIG:proj_dust_type}, currently
silicate and graphite grains are distributed in essentially the same manner.
While we may expect slight variations, given that AGB stars and SNe produce
dust of different compositions and the dust originating from SNe may be more
readily destroyed in SN shocks, future work should include
starlight-driven radiation pressure on dust grains.  Given that the dust
composition will impact the grain size distribution \citep{Mathis1977, Kim1994}
and radiation pressure is suggested to be important in simulations treating
dust \citep{Murray2005, Bekki2015b}, this physics may help differentiate
between silicate and graphite grains.
\end{enumerate}

While our current dust model is fairly passive and simplistic, it reproduces a
number of observed dust scaling relations surprisingly well at low redshift and
is a step towards a more complete treatment of dust in cosmological galaxy
formation simulations.  In the future, we can explore a live dust model that
includes dust-gas interactions more directly, enables dust cooling channels,
and allows for radiative forces to affect the motion of grains.  Together with
simulations of full cosmological volumes, this work will lead to a better
understanding of the full diversity of dust seen in the Universe.

\section*{Acknowledgements}

We thank Lia Corrales, Christopher Hayward, Alexander Ji, Rahul Kannan,
Federico Marinacci, Diego Mu{\~{n}}oz, and Gregory Snyder for helpful
discussion and Volker Springel for making \textsc{arepo} available.  We
are also grateful to the anonymous referee for comments that improved this
manuscript.

The simulations were performed on the joint MIT-Harvard computing cluster
supported by MKI and FAS.  RM acknowledges support from the DOE CSGF under
grant number DE-FG02-97ER25308.  MV acknowledges support through an MIT
RSC award.

\bibliographystyle{mn2e}
\bibliography{../bibliography}

\label{lastpage}

\end{document}